\documentclass[10pt]{article}
\usepackage{times}
\usepackage{epsf}
\newcommand{\figcap}[1]{
\it
\caption{#1}
  }
\newcommand{\dd}{\mbox{d}}

\newcommand{\bbox}[1]{\mbox{\boldmath $#1$}} 
\newcommand{\tfrac}[2]{{\textstyle\frac{#1}{#2}}}

\begin{document}
\begin{titlepage}
\title{On the Description \\ of \\ Quantum Vortices in Superfluid Films}
\author{J\"org P.\ Kottmann\thanks{Present address: Swiss Federal
Institute of Technology, Laboratory of EM Theory and Microwaves
Electronics, ETH-Zentrum, CH-8092 Zurich. E-mail:
kottmann@ifh.ee.ethz.ch} \hspace{.1cm} and Adriaan M.\ J.\
Schakel\thanks{E-mail: schakel@physik.fu-berlin.de} \\ Institut f\"ur
Theoretische Physik \\ Freie Universit\"at Berlin \\ Arnimallee 14,
14195 Berlin }
\date{\today}
\maketitle
\begin{abstract}
The dynamics of vortices in a superfluid film at the absolute zero of
temperature is studied.  The quantum-mechanical, i.e., ``first-quantized''
description is compared to a recently proposed quantum-field-theoretic,
i.e., ``second-quantized'' description.  The theory, consisting of the
nonrelativistic effective action of phonons linearly coupled to a
Chern-Simons term, is used to calculate the one-loop amplitude for
the elastic scattering of phonons from a vortex.  \vspace{.5cm} 
\end{abstract}
\end{titlepage}
\section{Introduction}
\label{sec:intro}
Vortices play an important role in several two-dimensional condensed
matter systems at low temperatures such as superfluids and
superconductors, fractional quantized Hall systems \cite{LeeFisher} and
Josephson junction arrays \cite{Stern}.  They are also believed to play
an important role in the quantum phase transitions these systems undergo
\cite{WenZe,MPAFisher,CFGWY}.  Two-dimensional vortices can ideally be
represented as pointlike and are, close to the absolute zero of
temperature, governed by the laws of quantum physics.  The quantum
mechanics of such pointlike vortices, which is known since long (see,
for example, Ref.\ \cite{YM}), is unusual.  This is because classically,
instead of obeying Newton's equations of motion, the coordinates of a
vortex center obey Kirchoff's equations of motion \cite{Lamb}.  As will
be recalled in Sec.\ \ref{sec:qm}, this is connected to a peculiar
canonical structure known to arise also in the problem of a charged
particle confined to move in a plane subject to an external
magnetic field \cite{Hamilton}.  As a result, the classical dynamics of
a vortex in a two-dimensional fluid resembles that of a charged particle
in a plane under the influence of a magnetic field \cite{Lamb}.

In recent years, various quantum-mechanical formulations of vortex
dynamics in superfluid films have been put forward which use this
analogy \cite{Popov,Hatsuda,Arovas}.  In these so-called dual approaches
(see Sec.\ \ref{sec:qm}), vortices are still described by delta
functions representing their worldlines, but their interaction is now
pictured as being mediated by a photon---the quantum of the
electromagnetic field---and not, as in the original formulation, by a
phonon---the quantum of the Goldstone field associated with the
spontaneously broken U(1) symmetry of a superfluid.  The analogy can be
understood \cite{OD} by comparing the continuity equation
\begin{equation} 
\partial_0 n + \nabla \cdot {\bf j} =0,
\end{equation} 
where $n$ is the particle number density and ${\bf j}$ the particle
number current of the superfluid, to the Maxwell equation (Gaussian
units will be used throughout):
\begin{equation} 
\frac{1}{c} \partial_0 H + \nabla \times {\bf E} = 0.
\end{equation} 
It is to be noted that in 2+1 dimensions, the magnetic field $H$ is a
scalar, and $\nabla \times {\bf E} = \epsilon_{ij} \partial_i E_j$, with
$\epsilon_{ij}$ ($i,j=1,2$) the antisymmetric Levi-Civita symbol
($\epsilon_{12} =1$).  It follows that with the replacements
\begin{equation} \label{corresp}
n \rightarrow H/\Phi_0, \hspace{.5cm} j_i \rightarrow \frac{e}{h}
\epsilon_{i j} E_j,
\end{equation} 
where $\Phi_0 = hc/e$ is the magnetic flux quantum, the continuity equation
goes over into the Maxwell equation.  The fundamental constants are included
in (\ref{corresp}) to match the dimensions on both sides of the
replacements.  Since the integral $\Phi(S) = \int_S \dd^2 x H$ represents
the magnetic flux penetrating the surface $S$, the ratio $H/\Phi_0$ is the
flux number density $n_\otimes$.  In other words, the duality transformation
maps the particles of the original formulation onto flux quanta in the dual
formulation.

In this paper, we take an entirely different approach.  Instead of
considering a quantum-mechanical, i.e., ``first-quantized'' description
of vortices in a superfluid film at the absolute zero of temperature, we
study a quantum-field-theoretic, i.e., ``second-quantized'' description
recently put forward by these authors \cite{JA}.  Vortices are here not
described by delta functions representing their worldlines as in the
quantum-mechanical approach, but by a nonsingular quantum field.  The
quantum field theory fulfills two stringent constraints.  The first one
arises because at zero temperature the entire system is superfluid, and
can, therefore, be considered an ideal fluid.  According to a theorem
due to Helmholtz \cite{Lamb}, a vortex in an ideal fluid moves with the
fluid, and, hence, has no independent dynamics.  The second constraint
stems from the following observation.  An electron traversing a closed
path in a magnetic field accumulates an Aharonov-Bohm phase given by ($2
\pi$ times) the number of flux quanta encircled \cite{AB}.  The above
analogy then asserts that an elementary vortex traversing a closed path
in a superfluid film also accumulates a geometrical phase given by ($2
\pi$ times) the number of particles encircled \cite{HW}.  It has been
shown in \cite{JA} (see also Sec.\ \ref{sec:qft}) that both these
constraints are satisfied by a specific Chern-Simons theory.  Two of the
main characteristics of a Chern-Simons theory in general are that,
first, it contains a vector field which has no independent dynamics, and
that, second, it imparts flux to particles.  By representing vortices by
a Chern-Simons field, we assured that they have no independent dynamics,
and that they see the particles as sources of geometrical phase.  To
connect with the standard quantum-mechanical approach, we introduce in
Sec.\ \ref{sec:qft} external vortices in the field theory and show that,
in this way, the known results are reproduced.

The specific Chern-Simons theory proposed in \cite{JA} is based on the
effective theory of phonons describing a superfluid film at low energies
and small momenta (see Sec.\ \ref{sec:eff}).  The effective Lagrangian
is the most general one consistent with the symmetries of the problem,
in particular Galilei invariance and invariance under global U(1)
transformations.  In Sec.\ \ref{sec:scat}, we use this effective theory
extended by the Chern-Simons part, representing the vortices, to
investigate one-loop contributions to the elastic scattering of phonons
by a vortex.  We apply dimensional analysis to identify the relevant
Feynman graphs.  To the order in which we are working, we do not
encounter ultraviolet divergences in the effective theory; only a
logarithmic infrared divergence arises.  This introduces an arbitrary
scale factor into the problem.  Often such a factor appears instead when
eliminating ultraviolet divergences by a redefinition of certain
parameters of the effective theory.  The one-loop contributions
evaluated in this paper give the first correction to an old result for
the scattering of a sound wave from a vortex obtained by Pitaevskii in
the Born approximation \cite{Pitaevskii}.

{\bf Notation} In the main text we will keep track of the fundamental
constants.  We will follow standard notation and use ${\bf p}$ as momentum
variable and ${\bf k}$ as wave-vector variable; they are related via ${\bf
p} = \hbar {\bf k}$, where $\hbar = h/2 \pi$ with $h$ Planck's constant.  The
(angular) frequency will be denoted by $\omega$ and the corresponding energy
by $p_0$; $p_0 = \hbar \omega$.  A spacetime point will be indicated as $x =
x_\mu = (t,{\bf x})$, $\mu = 0, 1, \ldots,d$, with $d$ the number of space
dimensions, the energy-momentum as $p = p_\mu = (p_0,{\bf p})$, and $k =
k_\mu = (\omega,{\bf k})$.
\section{Effective Theory for Sound Waves}
\label{sec:eff}
In this section, we briefly recall the nonrelativistic effective action
describing sound waves at low energy and small momentum in two space
dimensions \cite{effbos}.  As starting point to describe the system, we take
the microscopic model \cite{Beliaev} 
\begin{equation} \label{eff:Lagr}
{\cal L} = \phi^* \bigl[i \hbar \partial_0 - \epsilon(-i \hbar \nabla) +
\mu_0\bigr] \phi - \lambda_0 |\phi|^4,
\end{equation} 
where the complex scalar field $\phi(x)$ describes the atoms of mass $m$
constituting the liquid, $\epsilon(-i \hbar \nabla) = - \hbar^2 \nabla^2/2m$
is the kinetic energy operator, and $\mu_0$ is the chemical potential.  The
last term with a positive coupling constant, $\lambda_0 > 0$, represents a
repulsive contact interaction.

The Lagrangian has a  global U(1) symmetry, under which the matter field
transforms as 
\begin{equation} 
\phi(x) \rightarrow \phi'(x) = {\rm e }^{i \alpha}
\phi(x),
\end{equation}  
with $\alpha$ the transformation parameter.  At zero temperature, this
global symmetry is spontaneously broken by a nontrivial ground state.  This
can be easily seen by considering the potential energy
\begin{equation} \label{eff:V}
{\cal V} = - \mu_0 |\phi|^2 + \lambda_0 |\phi|^4,
\end{equation} 
which has, for $\mu_0 >0$, a minimum away from the origin $\phi = 0$ at
\begin{equation}  \label{eff:min}
|\bar{\phi}|^2 = \frac{1}{2} \frac{\mu_0}{\lambda_0 }.                     
\end{equation}
Since the total particle number density $n(x)$ is given by
\begin{equation} 
n(x) = |\phi(x)|^2,
\end{equation} 
the quantity $\bar{n}_0 := |\bar{\phi}|^2$ physically represents the number
density of particles contained in the condensate.

To account for the nontrivial ground state, we shift $\phi$ by the (complex)
constant $\bar{\phi}$ and write \cite{Popov}
\begin{equation}  \label{eff:newfields}
\phi(x) = {\rm e}^{i \varphi(x)} \, [\bar{\phi} + \tilde{\phi}(x)].
\end{equation}
The scalar field $\varphi(x)$ is a background field representing the
Goldstone mode of the spontaneously broken global U(1) symmetry.  In terms
of the new variables, the quadratic terms of the Lagrangian (\ref{eff:Lagr})
may be cast in the matrix form 
\begin{equation}  \label{eff:L0}
{\cal L}_0 = \frac{1}{2} \tilde{\Phi}^{\dagger} M_0
\tilde{\Phi}, \;\;\;\;\;\;  \tilde{\Phi} = \left(\begin{array}{l} \tilde{\phi} \\
\tilde{\phi}^* \end{array} \right),
\end{equation}
with
\begin{equation}   \label{eff:M} 
M_0 =   \left( \begin{array}{cc}
i\hbar \partial_0 - \epsilon + \mu_0 - U - 4 \lambda_0 |\bar{\phi}|^2 &
\!\!\!\! - 2 \lambda_0 \bar{\phi}^2 \\ - 2 \lambda_0 \bar{\phi}^*\mbox{}^2 &
\!\!\!\! -i\hbar \partial_0 - \epsilon + \mu_0 - U - 4 \lambda_0
|\bar{\phi}|^2
\end{array} \right),
\end{equation} 
and $U$ is the Galilei-invariant combination \cite{Popov}
\begin{equation}  \label{eff:U}
U = \hbar \partial_0 \varphi + \frac{1}{2m} (\hbar \nabla \varphi)^2.
\end{equation} 
In writing (\ref{eff:M}), we have omitted a term of the form $\nabla^2
\varphi$ containing two derivatives which is irrelevant in the regime of low
momentum in which we are interested.  We have also omitted a term of the
form $\nabla \varphi \cdot {\bf j}$, where ${\bf j}$ is the Noether current
associated with the global U(1) symmetry,
\begin{equation} 
{\bf j} =  \frac{\hbar}{2 i m}  (\phi^* \nabla \phi - \nabla  \phi^* \phi).
\end{equation}  
This term, which after a partial
integration becomes $- \varphi \nabla \cdot {\bf j}$, is also irrelevant at
low energy and small momentum because in a first approximation the particle
number density is constant, so that the classical current satisfies the
condition
\begin{equation} 
\nabla \cdot {\bf j} =0.
\end{equation}
The spectrum $E({\bf p})$ obtained from the matrix $M_0$ with the background
field $U$ set to zero is the famous single-particle Bogoliubov
spectrum \cite{Bogoliubov},
\begin{eqnarray}  \label{eff:bogo}
E({\bf p}) &=& \sqrt{ \epsilon ^2({\bf p}) + 2 \mu_0 \epsilon({\bf p}) }
\nonumber \\ &=& \sqrt{ \epsilon ^2({\bf p}) + 4 \lambda_0 |\bar{\phi}|^2
\epsilon({\bf p}) }. 
\end{eqnarray} 
The most notable feature of this spectrum is that it is gapless,
behaving for small momentum as
\begin{equation} \label{eff:micror}
E({\bf p}) \sim u_0 \, |{\bf p}|, 				
\end{equation} 
with $u_0 = \sqrt{\mu_0/m}$ a velocity which is sometimes referred to as the
microscopic sound velocity.  It was first shown by Beliaev \cite{Beliaev}
that (in three dimensions) the gaplessness of the single-particle spectrum
persists at the one-loop order.  This was subsequently proven to hold to all
orders in perturbation theory by Hugenholtz and Pines \cite{HP}.  For large
momentum, the Bogoliubov spectrum takes a form
\begin{equation} \label{eff:med}
E({\bf p}) \sim \epsilon({\bf p}) + 2 \lambda_0 |\bar{\phi}|^2
\end{equation} 
typical for a nonrelativistic particle with mass $m$ moving in a medium.  To
highlight the condensate, we have chosen here the second equality in
(\ref{eff:bogo}), where $\mu_0$ is replaced with $2 \lambda_0 |\bar{\phi}|^2$.

The gaplessness of the single-particle spectrum (\ref{eff:bogo}) is a result
of Goldstone's theorem.  This can, for example, be seen by considering the
relativistic version of the theory.  There, one finds two spectra: one
corresponding to a massive Higgs particle which in the nonrelativistic limit
becomes extremely heavy and decouples from the theory, and one corresponding
to the Goldstone mode of the spontaneously broken global U(1) symmetry
\cite{BBD}.  The latter reduces in the nonrelativistic limit to the
Bogoliubov spectrum.  Another way to see this is to couple the theory
(\ref{eff:Lagr}) to an electromagnetic field.  The single-particle spectrum
then acquires an energy gap.  This is what one expects to happen to the
spectrum of a Goldstone mode when the Higgs mechanism is operating.  In
other words, the single-particle spectrum (\ref{eff:bogo}) is identical to
that of the Goldstone mode which physically represents the collective
density fluctuations in the superfluid.  The equivalence of the
single-particle excitation and the collective density fluctuation has been
proven to all orders in perturbation by Gavoret and Nozi\`eres \cite{GN}.

\begin{figure}
\begin{center}
\epsfxsize=8.cm
\mbox{\epsfbox{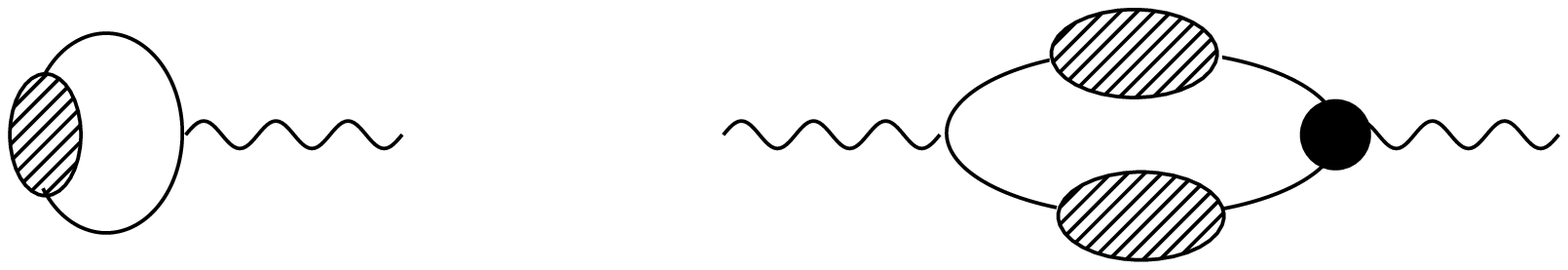}}
\end{center}
\figcap{Graphical representation of the effective theory
(\protect\ref{eff:Leff}). The symbols are explained in the
text. \label{fig:effective}}
\end{figure}

To find the effective theory governing the superfluid at low energy and
small momentum from (\ref{eff:L0}), we integrate out the fluctuating field
$\tilde{\Phi}$ \cite{effbos}.  The first two terms of the effective theory
are graphically represented by Fig.\ \ref{fig:effective}.
A line with a shaded bubble inserted stands for $i\hbar$ times the {\it
full} Green function $G$ and the black bubble denotes $i/\hbar$ times the
{\it full} interaction $\Gamma$ of the $\tilde{\Phi}$-field with the
background field $U$ which is denoted by a wiggly line.  Both $G$ and
$\Gamma$ are $2 \times 2$ matrices.  The full interaction is obtained from
the inverse Green function by differentiation with respect to the chemical
potential,
\begin{equation}  \label{bcs:defga}
\Gamma = - \frac{\partial G^{-1}}{\partial \mu}.		
\end{equation}
This follows because $U$, as defined in (\ref{eff:U}), appears in the theory
only in the combination $\mu - U$.  To lowest order, the inverse Green
function is given by the matrix $M_0$ in (\ref{eff:M}), so that the vertex
describing the interaction between the $\tilde{\Phi}$ and $U$-fields is
$i/\hbar$ times minus the unit matrix.  Because in terms of the full Green
function $G$, the particle number density reads
\begin{equation}
\bar{n} = \frac{i \hbar }{2} \, {\rm tr} \int \frac{\dd^{d+1} p}{h^{d+1}} G
(p),
\end{equation}
where $\dd^{d+1} = \dd p_0 \dd^d p$, it follows that the first diagram in
Fig.\ \ref{fig:effective} stands for $-\bar{n} U$.  The bar over $n$ is to
show that the particle number density obtained in this way is a constant,
representing the density of the uniform system with $U(x)$ set to zero.  The
second diagram without the wiggly lines denotes $i/\hbar$ times the (0
0)-component of the {\it full} polarization tensor, $\Pi_{0 0}$, at zero
energy transfer and low momentum ${\bf q}$,
\begin{equation} \label{eff:pi}
\frac{i}{\hbar} \lim_{{\bf q} \rightarrow 0} \Pi_{0 0}(0,{\bf q}) =
-\frac{1}{2} \lim_{{\bf q} \rightarrow 0} {\rm tr} \int \frac{\dd^{d+1}
p}{h^{d+1}} G \, \Gamma \, G \, (p_0,{\bf p}+ {\bf q}).
\end{equation} 
The factor $\tfrac{1}{2}$ is a symmetry factor which arises because the two
Bose lines are identical.  We proceed by invoking an argument due to Gavoret
and Nozi\`eres \cite{GN} to relate the left-hand side of (\ref{eff:pi}) to
the sound velocity.  By virtue of relation (\ref{bcs:defga}) between the
full Green function $G$ and the full interaction $\Gamma$, the (0
0)-component of the polarization tensor can be cast in the form
\begin{eqnarray}  \label{bcs:cruc}
\lim_{{\bf q} \rightarrow 0} \Pi_{0 0} (0,{\bf q}) &=& - \frac{i \hbar}{2}
\lim_{{\bf q} \rightarrow 0} {\rm tr} \int \frac{\dd^{d+1} p}{h^{d+1}} G \,
\frac{\partial G^{-1}}{\partial \mu} \, G (p_0,{\bf p}+ {\bf q}) \nonumber
\\ &=& \frac{i \hbar}{2} \frac{\partial }{\partial \mu} \lim_{{\bf q}
\rightarrow 0} {\rm tr} \int \frac{\dd^{d+1} p}{h^{d+1}} G (p_0,{\bf p}+ {\bf q})
\nonumber \\ &=& \frac{\partial \bar{n}}{\partial \mu} = - \frac{1}{V}
\frac{\partial \Omega}{\partial \mu^2},
\end{eqnarray} 
where $\Omega$ is the thermodynamic potential and $V$ the volume of the
system.  The right-hand side of (\ref{bcs:cruc}) is $\bar{n}^2 \kappa$, with
$\kappa$ the compressibility.  Because it is related to the macroscopic
sound velocity $c$ via
\begin{equation}
\kappa = \frac{1}{m \bar{n} c^2},
\end{equation}
we conclude that the (0 0)-component of the full polarization tensor
satisfies the so-called compressibility sum rule of statistical
physics \cite{GN} 
\begin{equation}          \label{bec:rel}           
\lim_{{\bf q} \rightarrow 0} \Pi_{0 0} (0,{\bf q}) = \frac{\bar{n}}{m c^2}.
\end{equation}
Putting the pieces together, we infer that the diagrams in Fig.\
\ref{fig:effective} stand for the effective theory \cite{effbos}
\begin{equation} \label{eff:Leff}  
{\cal L}_{{\rm eff}} = -\bar{n}\left[\hbar\partial_{0}\varphi +
\frac{1}{2m}(\hbar {\bf \nabla} \varphi)^{2} \right] + \frac{\bar{n}}{2m
c^{2}}\left[\hbar\partial_{0}\varphi + \frac{1}{2m}(\hbar {\bf
\nabla}\varphi)^{2}\right]^{2},
\end{equation} 
where we recall that ${\bar n}$ is the particle number density of the fluid
at rest.  This nonlinear theory describes a nonrelativistic sound wave, with
the dimensionless real scalar field $\varphi$ representing the Goldstone
mode of the spontaneously broken global U(1) symmetry.  It has the gapless
dispersion relation $\omega^2({\bf k}) = c^2 {\bf k}^2$, where $\omega$ is
the (angular) frequency, ${\bf k}$ the wave vector, and $c$ the sound
velocity.  This effective theory gives a complete description of the
superfluid valid at low energies and small momenta.  There are of course
higher-order terms, but they need to be included only as one goes to higher
energies and momenta.  The same effective theory we discussed here appears
also in the context of (neutral) superconductors and that of classical
hydrodynamics \cite{eff}.

The form of the effective theory (\ref{eff:Leff}) can also be derived
from general symmetry arguments \cite{Takahashi,GWW}.  The basic idea
is that the presence of a gapless Goldstone has to be reconciled with
Galilei invariance, which demands that the mass current and the momentum
density are equal.  This leads to the conclusion that the U(1) Goldstone
field $\varphi$ can only appear in the combination (\ref{eff:U}).  To
obtain the required linear spectrum for the Goldstone mode it is
necessary then to have the form (\ref{eff:Leff}).

The particle number current that follows from (\ref{eff:Leff}) reads
\begin{eqnarray}  
n(x) &=& \bar{n} -\frac{\bar{n}}{m c^{2}} \left\{\hbar \partial_{0}
\varphi(x) + \frac{1}{2 m} [\hbar {\bf \nabla} \varphi(x)]^{2}\right\}
\label{roh1} \\ {\bf j}(x) &=& n(x) {\bf v}(x), \label{roh2}
\end{eqnarray} 
where ${\bf v} = (\hbar/m) \nabla \varphi$ is the superfluid velocity field.
Equation (\ref{roh1}) reflects Bernoulli's principle which states that in
regions of rapid flow, the density and therefore the pressure is low.

The diagrams of Fig.~\ref{fig:effective} can also be evaluated in a loop
expansion to obtain explicit expressions for the particle number density
$\bar{n}$ and the sound velocity $c$ to a given order \cite{effbos}.  In
doing so, one encounters---apart from ultraviolet divergences which can be
eliminated by renormalization---also infrared divergences because the
Bogoliubov spectrum is gapless.  When, however, all one-loop contributions
are added together, these divergences are seen to cancel \cite{effbos}.  One
finds for $d=2$ to the one-loop order
\begin{equation} \label{eff:nc}
\bar{n} =  \frac{1}{2} \frac{\mu}{\lambda}, \;\;\;\;
c^2 =  \frac{\mu}{m} = 2 \frac{\lambda}{m} \bar{n},
\end{equation} 
where $\lambda$ and $\mu$ are the renormalized coupling constants (see
below).  The second expression in (\ref{eff:nc}) for the sound velocity
is appropriate when, as is often the case in experiment, the particle
number is fixed. The first one, featuring the chemical potential, is
appropriate in the presence of a reservoir with which the system can
freely exchange particles, so that only the {\it average} particle
number is fixed.

Given the form of the effective theory, the particle number density and
sound velocity can be more easily obtained directly from the
thermodynamic potential $\Omega$ via
\begin{equation}  \label{bec:thermo}
\bar{n} = - \frac{1}{V} \frac{\partial \Omega }{\partial \mu}; \;\;\;\;\;\;
\frac{1}{c^2} = - \frac{1}{V} \frac{m}{\bar{n}} \frac{\partial^2 \Omega
}{\partial \mu^2},
\end{equation} 
where $V$ is the volume of the system.  In this approach, one only has to
calculate the thermodynamic potential which at zero temperature and in the
approximation we are working is given by the effective potential ${\cal
V}_{\rm eff}$ corresponding to the theory (\ref{eff:Lagr}): $\Omega = \int
\dd^d x {\cal V}_{\rm eff}$.  To the one-loop order, the effective potential
for a uniform system reads
\begin{equation} 
{\cal V}_{\rm eff} = - \frac{\mu^2_0}{4 \lambda_0} + \frac{1}{2} \int
\frac{\dd^d p}{h^d} E({\bf p}),
\end{equation} 
with $E({\bf p})$ the gapless Bogoliubov spectrum (\ref{eff:bogo}).  The
integral over the loop momentum in arbitrary space dimension $d$ yields
\begin{equation} \label{Veff}
{\cal V}_{\rm eff} = - \frac{\mu_0^2}{4 \lambda_0} - \frac{\Gamma(1-d/2)
\Gamma(d/2 + 1/2)}{2 \pi^{(d + 1)/2} \hbar^d \Gamma(d/2+2)} m^{d/2}
\mu_0^{d/2 + 1},
\end{equation}  
where we used the integral representation of the Gamma function
\begin{equation}  \label{gamma}
\frac{1}{a^z} = \frac{1}{\Gamma(z)} \int\limits_0^\infty \frac{\dd \tau}{\tau}
\tau^z {\rm e}^{-a \tau}
\end{equation}
together with dimensional regularization to suppress irrelevant
ultraviolet divergences.  By this we mean contributions which, for
regularization with a momentum cutoff $\Lambda$, would diverge with a
strictly positive power of $\Lambda$.  Note that we, following Ref.\
\cite{NP}, first carried out the integrals over the loop energies, and
then analytically continued the remaining integrals over the loop
momenta to arbitrary space dimensions $d$.

Expanded around $d=2$, Eq.\ (\ref{Veff}) gives
\begin{equation} \label{epsilon}
{\cal V}_{\rm eff} = - \frac{\mu^2_0}{4 \lambda_0} - \frac{1}{4 \pi \hbar^2
\epsilon} \frac{m \mu^2_0}{\kappa^\epsilon} + {\cal O}(\epsilon^0),
\end{equation} 
where $\epsilon = 2-d$ is the deviation from the upper critical
dimension $d=2$, and $\kappa$ is an arbitrary renormalization group
scale parameter, with the dimension of an inverse length.  The
right-hand side of (\ref{epsilon}) is seen to diverge in the limit $d
\rightarrow 2$.  The theory can be rendered ultraviolet finite by
introducing a renormalized coupling constant $\lambda$ via
\begin{equation} \label{eff:lambdar}
\frac{1}{\lambda_0} = \frac{1}{\kappa^\epsilon}
\left(\frac{1}{\hat{\lambda}} - \frac{m}{\pi \hbar^2 \epsilon}\right),
\end{equation}  
where $\hat{\lambda} = \lambda/ \kappa^\epsilon$.  Its definition is
such that for arbitrary $d$, $\hat{\lambda}$ has the same engineering
dimension as $\lambda_0$ in the upper critical dimension $d=2$.  As
renormalization prescription we used the modified minimal subtraction.
The beta function $\beta(\hat{\lambda})$ follows as \cite{Uzunov}
\begin{equation} 
\beta(\hat{\lambda}) = \kappa \left. \frac{\partial \hat{\lambda}}{\partial
\kappa} \right|_{\lambda_0} = -\epsilon \hat{\lambda} + \frac{m}{\pi \hbar^2}
\hat{\lambda}^2.
\end{equation}  
In the upper critical dimension, this yields only one fixed point, viz.\ the
infrared-stable fixed point $\hat{\lambda}^* = 0$.  Below $d=2$, this point
is shifted to $\hat{\lambda}^* = \epsilon \pi \hbar^2/m$.

It follows from (\ref{epsilon}) that the chemical potential is not
renormalized to this order.  Incidentally, from the view point of
renormalization, the mass $m$ is an irrelevant parameter in
nonrelativistic theories and can be scaled away.

The most remarkable aspect of the effective theory (\ref{eff:Lagr}) is
that it is nonlinear.  The nonlinearity is necessary to provide a
Galilei-invariant description of a gapless mode in a nonrelativistic
system.  Since the Goldstone field in (\ref{eff:Lagr}) is always
accompanied by a derivative, we see that  the nonlinear terms carry
additional factors of $\hbar |{\bf k}|/mc$, with $|{\bf k}|$ the wave
number.  They can therefore be ignored provided the wave number is
smaller than the inverse coherence length $\xi = \hbar/mc$,
\begin{equation}  \label{zwd}
|{\bf k}| < 1/\xi.
\end{equation} 
For example, for $^4$He the coherence length, or Compton wavelength, is
about 10 nm.  In this system, the bound (\ref{zwd}), below which the
nonlinear terms can be neglected, coincide with the region where the
spectrum is linear and the description in terms of a sound mode is
applicable.

We close this section discussing the apparent mismatch in the number of
degrees of freedom in the normal and superfluid state.  Whereas the
normal state is described by a complex field $\phi$, the superfluid
state is described by just a single real scalar field $\varphi$.  The
resolution of this paradox lies in the spectrum of the modes
\cite{Leutwyler}.  In the normal state, the spectrum $E({\bf p}) = {\bf
p}^2/2m$ is linear in $E$, so that only positive energies appear in the
Fourier decomposition of the field $\phi$.  One needs therefore a
complex field to describe a single particle, as is well known from
standard quantum mechanics.  In the superfluid state, where the
spectrum $E^2({\bf p}) = c^2 {\bf p}^2$, is quadratic in $E$, the
Fourier decomposition of the field $\varphi$ contains positive as well
as negative energies.  As a result, a single real field suffices to
describe this mode.  In other words, although the number of fields is
different, the number of degrees of freedom is the same in the normal
and superfluid state.
\section{Quantum Mechanics of Vortices}
\label{sec:qm}
In this section, we discuss the conventional quantum-mechanical, or
``first-quantized'' description of vortices.  We will be working at the
absolute zero of temperature, so that the entire liquid is superfluid.
In the microscopic theory (\ref{eff:Lagr}), the asymptotic solution of a
static vortex with winding number $w$ centered at the origin is well
known \cite{Fetter}
\begin{equation}  \label{qm:sol}
\phi({\bf x}) = \sqrt{\frac{\mu}{2 \lambda}} \left(1 - \xi^2 \frac{w^2}{4
{\bf x}^2}\right) {\rm e}^{i w \theta} + {\cal O}\left(\frac{1}{{\bf x}^4}
\right),
\end{equation} 
where $\theta$ is the azimuthal angle and $\xi$ is the coherence length
introduced above (\ref{zwd}) which because of (\ref{eff:nc}) can also be
written as $\xi = \hbar/\sqrt{m\mu}$.  The density profile $n({\bf x})$ in
the presence of this vortex follows from taking $|\phi({\bf x})|^2$.

Let us now discuss how vortices can be incorporated in the effective theory.
To this end we follow Kleinert \cite{GFCM} and introduce a so-called plastic
field $\varphi_\mu^{\rm P} = (\varphi_0^{\rm P}, \bbox{\varphi}^{\rm P})$ in
the effective theory (\ref{eff:Leff}) via minimally coupling to the
Goldstone field:
\begin{equation} \label{hydro:minimal}
\tilde{\partial}_\mu \varphi \rightarrow \tilde{\partial}_\mu \varphi +
\varphi_\mu^{\rm P}, 
\end{equation}
with $\tilde{\partial}_\mu = (\partial_0,-\nabla)$.  If there are $N$
vortices with winding number $w_\alpha$ ($\alpha=1, \cdots, N$) and centered
at ${\bf X}^1(t), \cdots , {\bf X}^{N}(t)$, the plastic field satisfies the
relation
\begin{equation} \label{qm:pla}
\nabla \times \bbox{\varphi}^{\rm P}(x) = - 2 \pi \sum_\alpha w_\alpha
\delta[{\bf x} - {\bf X}^\alpha(t)],
\end{equation} 
so that we obtain for the velocity field 
\begin{equation} 
\nabla \times {\bf v} = \sum_\alpha \gamma_\alpha
\delta[{\bf x} - {\bf X}^\alpha(t)],
\end{equation}
as required.  Here, $\gamma_\alpha = (h/m) w_\alpha$, with $w_\alpha =
0, \pm 1, \pm 2, \ldots$, denotes the circulation of the $\alpha$th vortex
which is quantized in units of $h/m$.  A summation over vortex labels
will always be made explicit.  The combination $\tilde{\partial}_\mu
\varphi + \varphi_\mu^{\rm P}$ is invariant under the local gauge
transformation
\begin{equation} 
\varphi(x) \rightarrow \varphi(x) + \alpha(x); \;\;\;\;\;
\varphi^{\rm P}_\mu \rightarrow \varphi^{\rm P}_\mu - \tilde{\partial}_\mu
\alpha(x),
\end{equation} 
with $\varphi^{\rm P}_\mu$ playing the role of a gauge field.    

In the gauge $\varphi^{\rm P}_0=0$, a solution of Eq.\ (\ref{qm:pla}) is
given by
\begin{equation} \label{eff:pla}
\varphi^{\rm P}_i(x) = 2 \pi \epsilon_{ij} \sum_\alpha w_\alpha 
\delta_j[x,L_\alpha(t)] ,
\end{equation}
where $\epsilon_{ij}$ is the antisymmetric Levi-Civita symbol in two
dimensions, with $\epsilon_{12}=1$, and $\bbox{\delta} [x,L_\alpha(t)]$ is a
delta function on the line $L_\alpha(t)$ starting at the position ${\bf
X}^\alpha(t)$ of the $\alpha$th vortex and running to spatial infinity along
an arbitrary path:
\begin{equation} 
\delta_i [x,L_\alpha(t)] = \int_{L_\alpha(t)} \dd y_i \, \delta({\bf x} -
{\bf y}).
\end{equation} 

Let us for the moment concentrate on static vortices.  The field equation
obtained from the effective theory (\ref{eff:Leff}) with $\nabla \varphi$
replaced by the covariant derivative $\nabla \varphi - \bbox{\varphi}^{\rm
P}$ and $\partial_0 \varphi$ set to zero simply reads
\begin{equation}
\nabla \cdot {\bf v} = 0, \;\;\;\; {\rm or} \;\;\;\; \nabla \cdot (\nabla
\varphi - \bbox{\varphi}^{\rm P}) = 0,
\end{equation}
when the fourth-order term is neglected.  It can be easily solved to yield
\begin{equation} \label{qm:solution}
\varphi ({\bf x}) = - \int \dd^2 y \,  G({\bf x} - {\bf y}) \nabla \cdot
\bbox{\varphi}^{\rm P}({\bf y}),
\end{equation}
where $G({\bf x})$ is the Green function of the Laplace operator in two
space dimensions
\begin{equation} \label{green}
G({\bf x}) = \int \frac{\dd^2 k}{(2 \pi)^2} \frac{ {\rm e}^{i {\bf k} \cdot
{\bf x}}}{{\bf k}^2} = - \frac{1}{2 \pi} \ln( |{\bf x}|),
\end{equation}
i.e., $\nabla^2 G({\bf x}) = - \delta({\bf x})$.  For the velocity field
we obtain in this way the well-known expression \cite{Lamb}
\begin{eqnarray}  \label{qm:vortices}
v_i({\bf x}) &=& \epsilon_{ij} \sum_\alpha \gamma_\alpha \partial_j \int
\dd^2 y G({\bf x} - {\bf y}) \delta[{\bf y} - {\bf X}^\alpha(t)]
\nonumber \\ &=& - \frac{1}{2 \pi} \epsilon_{ij} \sum_\alpha
\gamma_\alpha \frac{x_j- X^\alpha_j}{| {\bf x}-{\bf X}^\alpha |^{2}} ,
\end{eqnarray} 
which is valid for ${\bf x}$ sufficiently far away from the vortex
cores.  Let us now specialize to the case of a single static vortex
centered at the origin.  On substituting the corresponding solution in
(\ref{roh1}), we find for the density profile in the presence of a
static vortex asymptotically
\begin{equation} 
n({\bf x}) = \bar{n} \left(1 - \xi^2 \frac{w^2}{2{\bf x}^2} \right).
\end{equation} 
Since this density profile is identical to the one obtained in the
microscopic theory [see Eq.\ (\ref{qm:sol})], this exemplifies that, with
the aid of plastic fields \cite{GFCM}, vortices are correctly
accounted for in the effective theory.

Let us proceed to investigate the dynamics of vortices in this formalism and
derive the action governing it.  We consider only the first part of the
effective theory (\ref{eff:Leff}).  In ignoring the higher-order terms, we
approximate the superfluid by an incompressible fluid for which the particle
number density is constant, $n(x) = \bar{n}$, see Eq.\ (\ref{roh1}).  This
is an approximation often used.  We again work in the gauge $\varphi^{\rm
P}_0=0$ and replace $\nabla \varphi$ by the covariant derivative $\nabla
\varphi - \bbox{\varphi}^{\rm P}$, with the plastic field given by
(\ref{qm:pla}).  The solution of the resulting field equation for $\varphi$
is again of the form (\ref{qm:solution}), but now it is time-dependent
because the plastic field is.  Substituting this in the action $S_{\rm eff}
= \int \dd t \, \dd^2 x {\cal L}_{\rm eff}$, we find after some
straightforward calculus
\begin{equation} \label{qm:action}
S_{\rm eff} = m \bar{n} \int \dd t\left[-\frac{1}{2} \sum_\alpha
\gamma_\alpha {\bf X}^\alpha \times \dot{\bf X}^\alpha + \frac{1}{2\pi}
\sum_{\alpha < \beta} \gamma_\alpha \gamma_\beta \ln(|{\bf X}^\alpha -
{\bf X}^\beta|) \right],
\end{equation} 
where a dot over a symbol denotes the time derivative.  This action yields
the well-known equations of motion for point vortices in an incompressible
two-dimensional superfluid \cite{Lamb,Lund}:
\begin{equation} 
\dot{X}_i^\beta(t) = - \frac{1}{2 \pi} \epsilon_{ij} \sum_{\alpha \neq
\beta} \gamma_\alpha \frac{X^\beta_j(t) - X^\alpha_j(t)}{| {\bf
X}^\beta(t)-{\bf X}^\alpha(t) | ^{2}} .
\end{equation} 
Note that $\dot{X}_i^\beta(t) = v_i\left[{\bf X}^\beta(t)\right]$, where
${\bf v}(x)$ is the superfluid velocity (\ref{qm:vortices}) with the
time-dependence of the vortex centers included.  This nicely illustrates the
result due to Helmholtz for ideal fluids \cite{Lamb} we alluded to in the
Introduction, that a vortex moves with the fluid, i.e., with the local
velocity produced by the other vortices in the system.  Experimental support
for this conclusion has been reported in Ref.\ \cite{YP}.

The action (\ref{qm:action}) leads to a twisted canonical structure.  To
display it, let us rewrite the first term of the corresponding Lagrangian
as
\begin{equation} 
L_1 = - m \bar{n} \sum_\alpha \gamma_\alpha X^\alpha_1 \dot{X}^\alpha_2,
\end{equation} 
where we ignored a total derivative.  It follows that the canonical
conjugate to the second component $X_2^\alpha$ of the center coordinate ${\bf
X}^\alpha$ is its first component \cite{YM}
\begin{equation} 
\frac{\partial L_1}{\partial \dot{X}_2^\alpha} = - m \bar{n} \gamma_\alpha
X^\alpha_1.
\end{equation} 
This implies that phase space coincides with real space and it gives the
commutation relation
\begin{equation} 
[X_1^\alpha, X_2^\beta ] = \frac{i}{w_\alpha} \ell^2 \delta^{\alpha
\beta},
\end{equation} 
where
\begin{equation} \label{qm:ell}
\ell = 1/\sqrt{2 \pi \bar{n}}
\end{equation} 
is a characteristic length.  Its definition is such that $2 \pi \ell^2$ is
the average area occupied by a particle of the superfluid film.  The
commutation relation leads to an uncertainty in the location of the vortex
centers
\begin{equation} 
\Delta X_1^\alpha \Delta X_2^\alpha \geq \frac{\ell^2}{2 |w_\alpha|} ,
\end{equation}   
which is inverse proportional to the particle number density.

Elementary quantum mechanics \cite{LL} tells us that in the quasi-classical
approximation to each unit cell (of area $h$) in phase space there
corresponds one quantum state.  That is, the total number of states is given
by
\begin{equation} 
\mbox{\# states} = \frac{1}{h} \int \dd p \, \dd q,
\end{equation} 
where $p$ and $q$ are a pair of canonically conjugate variables, and the
integral is over the entire phase space.  For a two-dimensional quantum
vortex with winding number $w_\beta$, this implies that the number of states
it can be in is
\begin{equation} 
\mbox{\# states}  = |w_\beta| \, \bar{n} S ,
\end{equation} 
where $S$ is the surface area of the sample.  In other words, every particle
in the superfluid film makes ($w_\beta$ times) an additional state
available to the vortex.

This phenomenon that phase space coincides with real space is known to
also arise in the Landau problem of a charged particle confined to move
in a plane subject to a (constant) background magnetic field.  There, it
leads to the well-known result that the number of states available
to the charged particle in each Landau level is  
\begin{equation} 
\mbox{\# states} = |e_\beta| \frac{H}{hc} S,
\end{equation} 
where $H$ is the magnetic field component perpendicular to the plane, and
\begin{equation} 
e_\beta = v_\beta e_0,
\end{equation}  
($v_\beta = 0, \pm 1, \pm 2, \ldots$) is the electric charge of the
particle given as a multiple of the unit of charge $e_0 (>0)$.  In terms of
the magnetic flux quantum $\Phi_0 = hc/e_0$, the number of states can be
rewritten as
\begin{equation} 
\mbox{\# states} = |v_\beta| \frac{H}{\Phi_0} S = |v_\beta|
\bar{n}_\otimes S,
\end{equation} 
where $\bar{n}_\otimes$ is the flux number density.  Hence, whereas the
number of states for vortices in a superfluid film is determined by the
particle number, here it is determined by the flux number.  This 
agrees with the replacement (\ref{corresp}) discussed in the
Introduction.

\begin{figure}
\begin{center}
\epsfxsize=12.cm
\mbox{\epsfbox{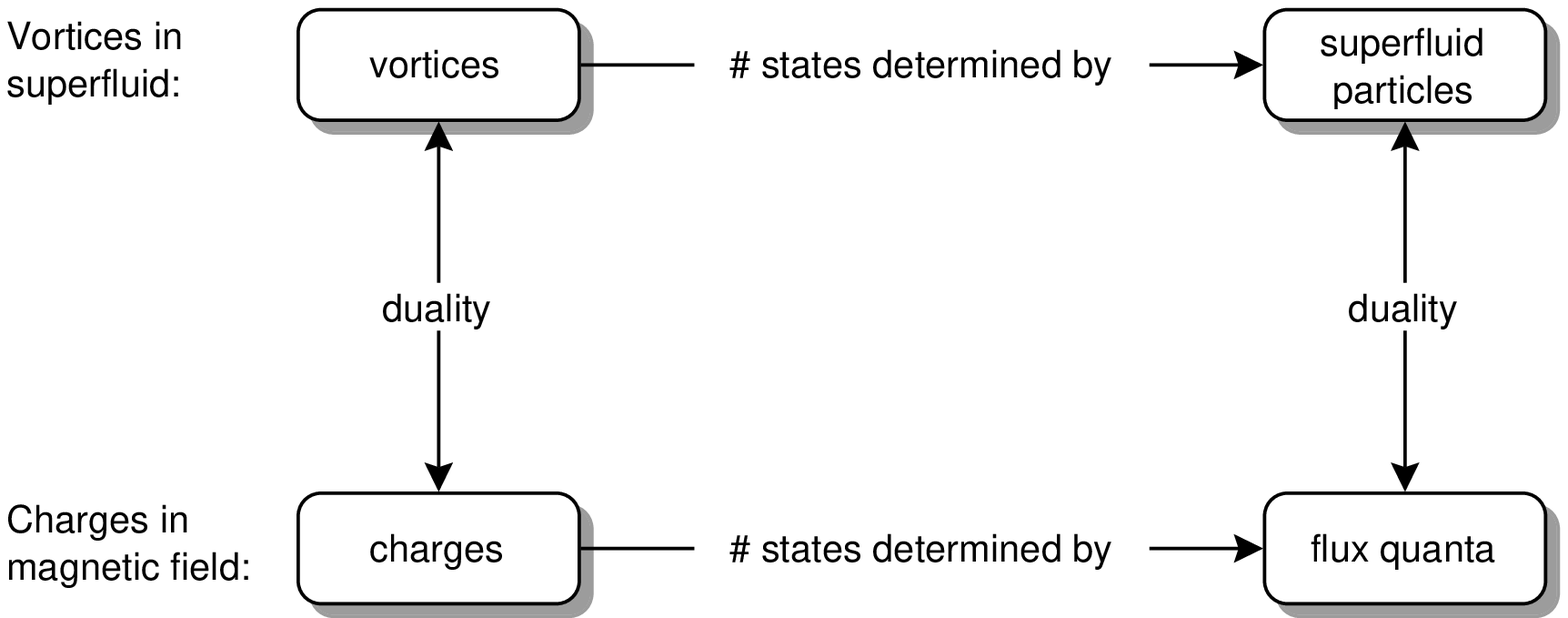}}
\end{center}
\figcap{Duality transformation. \label{fig:duality}}
\end{figure}

As remarked there, the analogy between the two problems is the basis of a
much used duality transformation.  If we collectively refer to the electric
charges in the Landau problem and the particles in the superfluid as
`charges', and to the vortices in the superfluid and the flux quanta in the
Landau problem as `vortices', then the duality transformation interchanges
charges and vortices (see Fig.\ \ref{fig:duality}).
Using this analogy, we see that the characteristic length (\ref{qm:ell})
translates into
\begin{equation} 
\ell_H = 1/\sqrt{2 \pi \bar{n}_\otimes} ,
\end{equation} 
which is precisely the magnetic length of the Landau problem.

We next turn to the geometrical phase \cite{Berry}.  In the
two-dimensional Landau problem, when a charged particle, say the $\beta$th,
is moved adiabatically around a close path $\Gamma_\beta$, its wavefunction
accumulates an extra Aharonov-Bohm phase factor $\gamma(\Gamma_\beta)$ given by
the Wilson loop:
\begin{equation} \label{qm:Berry}
W(\Gamma_\beta) = \exp\left[i \gamma(\Gamma_\beta) \right]
= \exp\left(\frac{i e_\beta}{\hbar c} \oint_{\Gamma_\beta} \dd {\bf x}
\cdot {\bf A}\right) = \exp \left[2 \pi i v_\beta \frac{H
S(\Gamma_\beta)}{\Phi_0}\right],
\end{equation} 
where ${\bf A}$ is the vector potential describing the background magnetic
field and $H S(\Gamma_\beta)$ the magnetic flux through the area
$S(\Gamma_\beta)$ spanned by the loop $\Gamma_\beta$.  The geometrical phase
$\gamma(\Gamma_\beta)$ in (\ref{qm:Berry}) is seen to be ($2 \pi v_\beta$
times) the number of flux quanta enclosed by the path $\Gamma_\beta$.

Because of the above analogy, it follows that the geometrical phase
acquired by the wavefunction of a vortex when it is moved adiabatically
around a closed path in the superfluid film is ($2 \pi w_\beta$ times) the
number of particles enclosed by the path \cite{HW}.
 
We have up to now considered a constant magnetic background field and an
{\it incompressible} superfluid characterized by a constant particle
number density.  A more general analysis of the geometrical phase for a
{\it compressible} superfluid was carried out by Haldane and Wu
\cite{HW}.  Their starting point was an {\it Ansatz} for the multivortex
wavefunction of the interacting Bose condensate \cite{Feynman}
\begin{equation} \label{qm:Ansatz}
\Psi = \prod_\alpha A({\bf X^\alpha}) f({\bf X}^\alpha) \Psi^0,
\end{equation} 
where $\Psi^0$ describes the condensate in the absence of vortices.  The
operator $A({\bf X})$ introduces an elementary vortex centered at ${\bf X}$,
\begin{equation} 
A({\bf X}) = \prod_a \left[ (x_1^a -X_1) + i (x_{2}^a - X_2)
\right],
\end{equation}
where ${\bf x}^a$ denotes the coordinate of the $a$th particle, and the
factor $f({\bf X})$ accounts for the change in the particle number density
owing to the presence of this vortex.  Repeating steps first carried out by
Arovas, Schrieffer, and Wilczek \cite{ASW} in the context of the fractional
quantized Hall effect, they cast the geometrical phase \cite{Berry},
\begin{equation}  
\gamma(\Gamma_\beta) = i \oint_{\Gamma_\beta} \dd {\bf X^\beta}
\cdot \langle \Psi |\nabla_{{\bf X}^\beta} \Psi \rangle, 
\end{equation} 
picked up by the wavefunction when the $\beta$th vortex is moved along
a closed path $\Gamma_\beta$ in the form
\begin{equation}   \label{Berry} 
\gamma (\Gamma_\beta) = \int \dd^2 x \oint_{\Gamma_\beta} \dd {\bf
X}^\beta \cdot \nabla \times \left[ \ln(|{\bf x}-{\bf
X}^\beta|) \right]  n({\bf x};{\bf X}^\alpha,{\bf X}^\beta) .
\end{equation}  
Here, $n({\bf x};{\bf X}^{\alpha},{\bf X}^{\beta})$ is the particle
number density in the presence of the transported vortex as well as the
other vortices centered at ${\bf X}^{\alpha}$ $(\alpha \neq \beta)$.
Writing
\begin{equation}  
n({\bf x};{\bf X}^\alpha,{\bf X}^\beta) = n({\bf x};{\bf X}^\alpha)
+ \delta n({\bf x}-{\bf X^{\beta}}) ,
\end{equation} 
where $n({\bf x};{\bf X}^\alpha)$ is the particle number density in the
absence of the vortex being moved, they concluded that, apart from
corrections owing to residual vortex interactions that become small in the
dilute-vortex limit, the geometrical phase is given by
\begin{equation}  \label{Hberry} 
\gamma (\Gamma_\beta) = 2\pi \int_{\Gamma_\beta} \dd^2 x \, n({\bf x};{\bf
X}^\alpha) .
\end{equation} 
That is, for a compressible superfluid a similar result is found as in the
simpler case of an incompressible superfluid, in that the geometrical phase
acquired by an elementary vortex being moved around a closed path is ($2 \pi$
times) the {\it average} number of particles enclosed by the path \cite{HW}.

The last topic we wish to discuss in this section on the quantum
mechanics of vortices is the elastic scattering of a sound wave from a vortex.
The subject, first studied by Pitaevskii \cite{Pitaevskii}, has recently
received considerable attention (see Ref.\ \cite{Sonin1} and references
therein).  To obtain the hydrodynamic equation governing this scattering
process, we first record the nonlinear field equation obtained from the
effective theory (\ref{eff:Leff}) \cite{eff}
\begin{equation} \label{qm:nonlinear}
\partial_0 \mu + c^2 \nabla \cdot {\bf v}
= \partial_0 {\bf v}^2 + \tfrac{1}{2} {\bf v} \cdot \nabla {\bf v}^2,
\end{equation} 
with $\mu = - (\hbar/m) \partial_0 \varphi$ the chemical potential per unit
mass.  If we ignore the nonlinear terms, Eq.\ (\ref{qm:nonlinear}) becomes
the more familiar wave equation
\begin{equation} \label{qm:wave}
\partial_0^2 \varphi - c^2 \nabla^2 \varphi = 0.
\end{equation}

Let us next introduce a freely moving, i.e., non-pinned vortex.  Driven by
the sound wave, the vortex will oscillate around some point, ${\bf x} = 0$
say.  The velocity field in the presence of a moving vortex is obtained from
the static solution [see Eq.\ (\ref{qm:vortices})]
\begin{equation} 
v_i({\bf x}) = \frac{1}{2 \pi} \gamma \epsilon_{ij}   \frac{x_j}{{\bf x}^2} ,
\end{equation} 
by replacing the coordinate ${\bf x}$ with ${\bf x} - {\bf v}_L t$, where
${\bf v}_L(t)$ is the velocity of the vortex.  This implies that
\begin{equation} \label{vchange}
\partial_0 {\bf v}_{\rm v}({\bf x} -{\bf v}_Lt) = -{\bf v}_L \cdot \nabla
{\bf v}_{\rm v}({\bf x} -{\bf v}_Lt),
\end{equation} 
where ${\bf v}$ is given an index v to show that it is the velocity field
produced by the vortex.  Since the solution ${\bf v}_{\rm v}$ is curl-free
outside the vortex core, the right-hand side of Eq.\ (\ref{vchange}) may be
written as $- \nabla ({\bf v}_L \cdot {\bf v}_{\rm v})$ there.  The
hydrodynamic equation governing the elastic scattering of a sound wave from
the free vortex now follows from writing the velocity field as
\begin{equation} 
{\bf v}(x) = {\bf v}_{\rm v}({\bf x} - {\bf v}_L t) + \frac{\hbar}{m} \nabla
\tilde{\varphi}(x),
\end{equation} 
with $\tilde{\varphi}$ describing small variations around the oscillating
vortex solution, and noting that (\ref{vchange}) requires that we write for
the chemical potential per unit mass
\begin{equation} \label{qm:mu}
\mu (x) = {\bf v}_L \cdot {\bf v}_{\rm v}({\bf x} - {\bf v}_L t) -
\frac{\hbar}{m} \partial_0 \tilde{\varphi} (x).
\end{equation}
Substituting this in the field equation (\ref{qm:nonlinear}), we obtain to
linear order in $\tilde{\varphi}$ the hydrodynamic equation \cite{Sonin1},
\begin{equation} \label{AB}
\partial^2_0 \tilde{\varphi}(x) - c^2 \nabla^2 \tilde{\varphi}(x) = - {\bf
v}_{\rm v}({\bf x}) \cdot \nabla \partial_0 [2 \tilde{\varphi}(x) -
\tilde{\varphi}(t,0)],
\end{equation}  
where we approximated ${\bf v}_{\rm v}({\bf x} - {\bf v}_L
t)$ by ${\bf v}_{\rm v}({\bf x})$.  To linear order in $\tilde{\varphi}$
this is allowed since the vortex, being driven by the sound wave, has a
velocity
\begin{equation} 
{\bf v}_L(t) = \frac{\hbar}{m} \nabla \tilde{\varphi}(t,{\bf x} = {\bf v}_L t)
\approx \frac{\hbar}{m} \nabla \tilde{\varphi}(t,0).
\end{equation}   
We also neglected a term quadratic in ${\bf v}_{\rm v}$ which is
justified because $|{\bf v}_{\rm v}| << c$ outside the vortex core
\cite{Pitaevskii}.  The first term at the right-hand side of (\ref{AB})
stems from the nonlinear term $\partial_0 {\bf v}^2$ in the field
equation (\ref{qm:nonlinear}).

Equation (\ref{AB}) has been used by Sonin \cite{Sonin1} as a basis to study
the elastic scattering of a sound wave from a free moving vortex.  In the
Born approximation, he found as two-dimensional scattering amplitude
$f(\theta,|{\bf k}|)$ the result first derived by Pitaevskii \cite{Pitaevskii}:
\begin{equation}  \label{pit}
f(\theta,|{\bf k}|) = \frac{1}{2} \sqrt{\frac{|{\bf k}|}{2 \pi}} \frac{h}{ m
c} {\rm e}^{i \frac{\pi}{4}} \frac{\sin{\theta}
\cos{\theta}}{1-\cos{\theta}} ,
\end{equation}  
where $\theta$ is the angle between the incoming and the scattered sound
wave, and $|{\bf k}|$ their wave number.

In the next section, we discuss a quantum field theory of vortices
recently proposed by these authors \cite{JA} and show that it
reproduces the results discussed here.
\section{Quantum Field Theory of Vortices}
\label{sec:qft}
In the previous section we have seen that vortices in a two-dimensional
superfluid at the absolute zero of temperature can be considered as
point particles which are subject to the laws of quantum physics.  It is
therefore natural to ask whether vortices also admit a
quantum-field-theoretic, or ``second-quantized'' description.  In a
recent letter, we have argued they indeed do and proposed the following
quantum field theory \cite{JA}:
\begin{equation} \label{vbn} 
{\cal L} = {\cal L}_{\rm eff} - e n a_{0} + \frac{e}{c} {\bf j} \cdot
{\bf a} + {\cal L}_{{\rm CS}}.
\end{equation} 
It consists of the nonrelativistic effective Lagrangian (\ref{eff:Leff})
describing the superfluid without vortices, {\it linearly} coupled via the
particle number current $j_\mu = (n,{\bf j})$ to a vector field $a_\mu
=(a_0,{\bf a})$ describing the vortices. This field is governed solely by a
Chern-Simons term
\begin{equation}
{\cal L}_{{\rm CS}} = \frac{1}{2 c} {\bf a} \times \partial_{0} {\bf a} -
a_{0} {\bf \nabla} \times {\bf a},
\end{equation} 
with the charge $e$ determined by \cite{JA}
\begin{equation}
e^2 = h c.
\end{equation}
Our convention is such that, in contrast to $j_\mu$, the dimensions of the
time and space components of the vector field $a_\mu$ are the same.  It is
important to note that the coupling of the sound mode to the Chern-Simons
field is linear and not minimal as in gauge theories.  The coupled theory
(\ref{vbn}) therefore does not possess a gauge invariance involving a
simultaneous local gauge transformation of the Chern-Simons field and the
matter field.  To appreciate why a minimal coupling is not feasible here,
note that this would inevitably result in the disappearance of the gapless
sound mode because of the Higgs mechanism.  This would disagree with the
physics we wish to describe, namely the dynamics of vortices in a {\it
compressible} superfluid film.  For the gapless sound mode to survive a
coupling to a gauge field, the coupling cannot be minimal so that there is
no local gauge invariance in the full theory.

The appearance of a Chern-Simons term in (\ref{vbn}) is expected for two
reasons.  First, as we discussed in the previous section, vortices in a
superfluid at the absolute zero of temperature always move with the fluid.
In other words, they have no independent dynamics.  This is precisely also a
basic property of a Chern-Simons term.  It comes about because the zeroth
component $a_0$ of the vector field is a Lagrange multiplier, demanding the
equality
\begin{equation} \label{para}
\nabla \times {\bf a} = - e n,
\end{equation} 
or when integrated
\begin{equation} \label{flux}
\Phi  = -e N,
\end{equation} 
where $N = \int \dd^2 x \, n$ is the particle number and $\Phi = \int \dd^2
x \nabla \times {\bf a}$ the ``magnetic'' flux associated with the
Chern-Simons field.  In the Coulomb gauge $\nabla \cdot {\bf a} = 0$, Eq.\
(\ref{para}) can be easily solved to yield
\begin{equation}  \label{aa}
{\bf a}(x) = -e \nabla \times \int
\dd^{2}y G({\bf x} - {\bf y}) n(t,{\bf y}) ,
\end{equation} 
where the Green function $G({\bf x})$ is given in (\ref{green}).  The
solution shows that the Chern-Simons field is entirely determined by the
particle number density $n$ and that it has, therefore, no independent
dynamics.

The second reason why a Chern-Simons term is expected to appear is that it
encodes the geometrical phase acquired by a vortex when it winds around a
boson.  The quantum-mechanical analog of such a term, representing the
linking number \cite{Kleinert} of a closed boson and vortex trajectory, was
introduced in the problem by Arovas and Freire \cite{Arovas}.  To calculate
the geometrical phase $\gamma(\Gamma)$ we again evaluate the Wilson loop
$W(\Gamma) = \exp[i \gamma (\Gamma)]$ obtained by integrating the
Chern-Simons field ${\bf a}$ around a closed path $\Gamma$ in the
superfluid, 
\begin{equation} \label{BerryCS}
\gamma(\Gamma) = \frac{e}{\hbar c} \oint_{\Gamma} \dd {\bf l}
\cdot {\bf a}(x) = \frac{e^2}{h c} \int \dd^{2}y \oint_{\Gamma}
\dd {\bf l} \cdot [ \nabla_{{\bf x}} \times \ln(|{\bf x} - {\bf y}|)
n(t,{\bf y}) ],
\end{equation} 
where we substituted the explicit form (\ref{aa}) for ${\bf a}$.  This
geometrical phase is precisely the one in (\ref{Berry}) obtained by Haldane
and Wu \cite{HW}.  Whereas in their quantum-mechanical description, the
vortex being taken around the closed path $\Gamma$ sees the encircled bosons
as sources of geometric phase, here this counting is provided by the flux
imparted to a particle by the Chern-Simons term [see Eq.\ (\ref{flux})].

We next introduce external point vortices into the theory to determine their
action and compare it with the action (\ref{qm:action}) obtained from the
quantum-mechanical description of vortices.  As we did there, we approximate
the superfluid by an incompressible fluid and consider only the following
terms of the Lagrangian (\ref{vbn}):
\begin{equation}  \label{rel}
{\cal L}_{\rm ext} = -\bar{n} \left[\hbar\partial_{0}\varphi +
\frac{1}{2m}(\hbar {\bf \nabla} \varphi)^{2} \right] + \frac{e}{c} {\bf
j}\cdot {\bf a}^{\rm ext}.
\end{equation} 
The particle number current ${\bf j}$ reads ${\bf j} = \bar{n} {\bf v}$
in this approximation, and the Chern-Simons field ${\bf a}^{\rm
ext}$ is given by (\ref{aa}) with $n$ replaced by the external vortex
density
\begin{equation} 
n_{\rm ext}(x) = \sum_\alpha w_\alpha \delta[{\bf x} - {\bf
X}^\alpha(t)].
\end{equation}
Explicitly,
\begin{equation}  \label{grada}
{\bf a}^{\rm ext} (x) = -\frac{e}{2 \pi} \sum_\alpha w_\alpha
\nabla \arctan\left( \frac{x_2 - X_2^\alpha(t)}{x_1 - X_1^\alpha(t)}
\right),
\end{equation} 
or
\begin{equation} \label{aex}
a^{\rm ext}_i(x) = \frac{e}{2 \pi} \epsilon_{ij} \sum_\alpha
w_\alpha \frac{x_j - X_j^\alpha(t)}{|{\bf x} - {\bf X}^\alpha(t)|^2},
\end{equation} 
where ${\bf X}^\alpha(t)$ denotes the center of the $\alpha$th vortex with
winding number $w_\alpha$.  Note that the gauge $\nabla \cdot {\bf a} = 0$
is satisfied by the solution (\ref{aex}) only for ${\bf x} \neq {\bf
X}^\alpha$, i.e., outside the vortex cores.  The field equation for
$\varphi$ derived from (\ref{rel}) can be easily solved to yield
\begin{equation} \label{feq}
\varphi(x) = - \frac{e}{\hbar c} \int \dd^{2}y G({\bf x} - {\bf
y}) \, \nabla \cdot {\bf a}^{\rm ext}(t,{\bf y}) .
\end{equation}
Upon taking the gradient of this equation, we obtain as velocity field ${\bf
v} = (\hbar/m) \nabla \varphi$ the required form (\ref{qm:vortices}).  Apart
from a prefactor this expression coincides with the one for ${\bf a}^{\rm
ext}$ in (\ref{aex}).  This is because ${\bf a}^{\rm ext}$ can be written as
a gradient of a scalar function [see (\ref{grada})].  When substituting the
solution (\ref{feq}) back into the Lagrangian (\ref{rel}), which is
tantamount to integrating out the field $\varphi$, we recover precisely the
action (\ref{qm:action}).  We thus see that the theory (\ref{rel}) with its
linear coupling to a Chern-Simons field correctly reproduces the
two-dimensional action of point vortices in an incompressible superfluid.

Let us continue by considering the lowest order elastic scattering
amplitude of two phonons calculated from the effective theory.  In the
frame where the sum of the momenta of the two incoming phonons is zero,
the class of diagrams involving the exchange of Chern-Simons quanta
represents the scattering of a phonon from a vortex.  An analogous
situation arises in the context of Aharonov-Bohm scattering, i.e.,
elastic scattering of a charged nonrelativistic particle from an
infinitely thin magnetic flux tube.  It was pointed out by Bergman and
Lozano \cite{BL} that such a scattering process can also be described by
a quantum field theory consisting of a nonrelativistic $|\psi|^4$-theory
(with zero chemical potential) coupled to a Chern-Simons term---albeit
minimally.

At small energies and momenta, we can neglect the higher-order terms in the
Lagrangian (\ref{vbn}) and restrict ourselves to terms at most quadratic in
the field $\varphi$:
\begin{eqnarray}   \label{theory}
{\cal L}^{(2)} = \frac{\bar{n} }{mc^2} \Biggl\{&&
\!\!\!\!\!\!\!\!\!\!\!\!  \frac{1}{2}(\hbar \partial_{0}\varphi)^2 -
\frac{c^{2}}{2} (\hbar \nabla \varphi)^2 + e a_{0} \left[\hbar
\partial_0 \varphi + \frac{1}{2m} (\hbar \nabla \varphi)^{2} \right]
\nonumber \\ && \!\!\!\!\!\!\!\!\!\! - \frac{e\hbar^2}{mc} \partial_{0}
\varphi \nabla \varphi \cdot {\bf a}\Biggr\} + {\cal L}_{\rm CS} .
\end{eqnarray}
If we again introduce external vortices by replacing the Chern-Simons
field ${\bf a}$ with (\ref{aex}), the field equation for
$\varphi(t,{\bf x} \neq {\bf X}^\alpha)$ becomes 
\begin{equation} \label{phi-feq}
\partial_0^2 \varphi - c^2 \nabla^2 \varphi = 2 \frac{e}{mc} \partial_0
\nabla \varphi \cdot {\bf a}^{\rm ext},
\end{equation} 
where, in accordance with the derivation of the hydrodynamic equation
(\ref{AB}), we omitted a term proportional to $\partial_0 {\bf a}^{\rm
ext}$.  Given the form (\ref{aex}) of ${\bf a}^{\rm ext}$, we recognize in
the field equation (\ref{phi-feq}) derived from our theory, the hydrodynamic
equation (\ref{AB}) with the contribution from the vortex motion ignored.
Since Eq.\ (\ref{AB}) was used as starting point for the quantum-mechanical
description of the elastic scattering of a sound wave from a vortex, we
expect the effective theory (\ref{theory}) to be the appropriate starting
point for the quantum-field-theoretic description.

The propagators and vertices needed to calculate the scattering amplitude
are readily obtained from that Lagrangian (\ref{theory}).  In the gauge
$\nabla \cdot {\bf a} = 0$, the nonzero components of the Chern-Simons
propagator follow as 
\begin{equation}   \label{CSprop}
\raisebox{-0.2cm}{\epsfxsize=3.cm \epsfbox{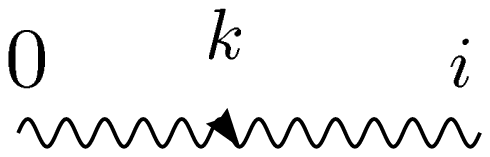} } \;\;\; : \;\;\; i
\hbar G_{i0}({\bf k}) = - i \hbar G_{0i}({\bf k}) = - \hbar
\epsilon_{ij} \frac{k_{j}}{{\bf k}^{2}},
\end{equation}  
where ${\bf k}$ is the wave vector.  Note that there is no frequency
dependence here.  The phonon propagator reads
\begin{equation}  \label{phiprop}
\raisebox{-0.2cm}{\epsfxsize=2.7cm \epsfbox{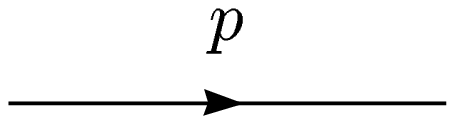} } \;\;\; : \;\;\; i
\hbar G(p) = \frac{m c^{2}}{\bar{n}} \frac{i
\hbar}{p_{0}^{2}-c^{2}{\bf p}^2 +i \eta} ,
\end{equation} 
where $\eta$ is a small positive constant that has to be taken to zero after
the integration over the loop energy $p_0$ has been carried out. In
(\ref{phiprop}), ${\bf p}$ denotes the momentum.  As vertices we read off
\begin{eqnarray}   \label{een} 
\raisebox{-0.2cm}{\epsfxsize=3.cm \epsfbox{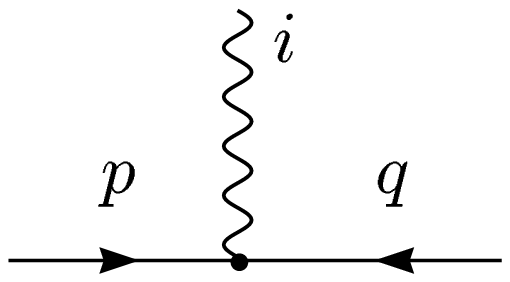} } &:& \frac{i}{\hbar}
\Gamma_i(p,q) = - \frac{i}{\hbar} \frac{e\bar{n}}{m^2 c^{3}}(p_{0} q_{i} +
p_{i} q_{0} ) \\ \label{zwee} \raisebox{-0.2cm}{\epsfxsize=3.cm
\epsfbox{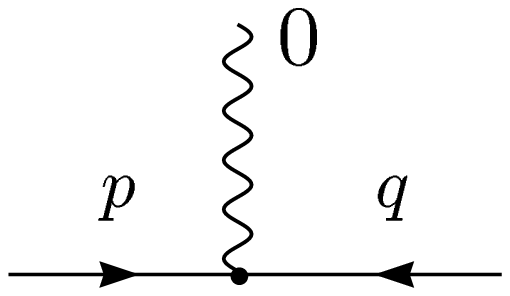} } &:& \frac{i}{\hbar} \Gamma_0(p,q) = - \frac{i}{\hbar}
\frac{e \bar{n}}{m^2 c^{2}} {\bf p} \cdot {\bf q} \\ && \nonumber \\
\label{dree}  
\raisebox{-0.2cm}{\epsfxsize=3.1cm \epsfbox{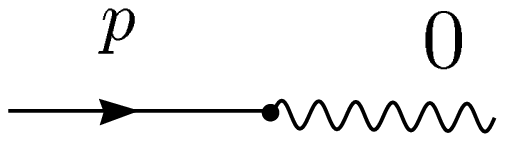} } &:& \frac{1}{\hbar}
\frac{e\bar{n}}{m c^{2}} p_{0}.
\end{eqnarray} 
In \cite{JA}, we calculated the scattering amplitude $A(\theta,|{\bf p}|)$  
to lowest order, with the result 
\begin{equation}    \label{treeres} 
\raisebox{-1.5cm}{\epsfxsize=3.cm \epsfbox{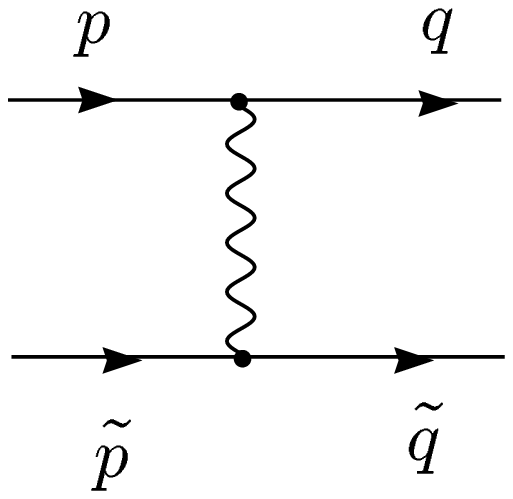} } \;\;\; : \;\;\; i
A^{(0)}(\theta,|{\bf p}|) = -\frac{1}{2} \frac{e^2 |{\bf p}|}{m^2 c^2}
\frac{\sin{\theta} \cos{\theta}}{1-\cos{\theta}},
\end{equation} 
where, as in (\ref{pit}), $\theta$ is the scattering angle, and $|{\bf p}|$
the momentum of the incoming and scattered phonons.  In deriving this, use
is made of the following relations between the energy-momenta of the
incoming and outgoing phonons
\begin{equation} \label{relations}
p_0 = \tilde{p}_0 = q_0 = \tilde{q}_0, \;\;\; |{\bf p}|=|\tilde {\bf p}|
= |{\bf q}| = |\tilde{\bf q}| = p_0/c, \;\;\; \tilde{\bf p} = - {\bf p},
\;\;\; \tilde{\bf q} = - {\bf q}, 
\end{equation} 
and
\begin{equation} 
{\bf p} \cdot {\bf q} = {\bf p}^2 \cos(\theta), 
\end{equation} 
in the frame defined by the condition that the sum of the momenta of the
incoming phonons be zero.  The normalization $\zeta$ of the external lines
was determined in \cite{JA} to be given by the dimensionless factor
\begin{equation}  \label{zeta}
\zeta = \sqrt{\frac{m c^2}{2 p_0 \bar{n} A}} .
\end{equation} 
where $A$ is the surface area of the system.  Apart from a kinematic
factor, which is due to a different definition of the scattering amplitude
in quantum mechanics and in quantum field theory, Eq.\ (\ref{treeres})
agrees with the result (\ref{pit}) of Pitaevskii \cite{Pitaevskii}.  Hence,
using the effective theory (\ref{theory}), we have reproduced the Born
approximation of the scattering of a sound wave from a vortex.

For identical particles, one must add to the diagram in (\ref{treeres}) the
crossed diagram with the two outgoing lines exchanged.  This yields the same
result as in (\ref{treeres}) with $\theta \rightarrow \theta - \pi$.  Adding
the two contributions, we find 
\begin{equation} \label{tot}
A^{(0)}_{\rm tot}(\theta,|{\bf p}|) = i \frac{h \alpha}{m} \cot(\theta),
\end{equation} 
where we introduced the dimensionless parameter 
\begin{equation} \label{alpha}
\alpha = \xi |{\bf k}|,
\end{equation} 
which according to (\ref{zwd}) was assumed to be small.  Remember that
${\bf p} = \hbar {\bf k}$.  The ratio $h/m$ in (\ref{tot}) is the
circulation quantum.  As observed by Sonin \cite{Sonin1}, the result
(\ref{tot}) is identical to the lowest-order Aharonov-Bohm scattering
amplitude, where the parameter $\alpha$ denotes the magnetic flux
through the flux tube measured in units of $\Phi_0$, and $m$ the mass of
the scattered particle.  The ratio $h/m$ should now be interpreted not
as the circulation quantum, but as the dispersion constant \cite{Fick}
which determines the dispersion of a matter wave with wave vector ${\bf
k}$ describing a nonrelativistic particle of mass $m$,
\begin{equation} 
\omega({\bf k}) = \frac{\hbar}{2 m} {\bf k}^2.
\end{equation} 
In the quantum-field-theoretic approach to Aharonov-Bohm scattering
\cite{BL}, the contribution (\ref{tot}), which diverges for both $\theta
= 0$ and $\pi$, is due to the exchange of a single Chern-Simons quantum.
Since the Aharonov-Bohm scattering amplitude is known exactly, the next
term in a perturbative calculation of this amplitude is known to be of
order $\alpha^3$.
\section{One-Loop Corrections} 
\label{sec:scat}  
In this section, we consider one-loop corrections to the tree amplitude
(\ref{tot}).  We are interested in the next order in an expansion in
powers of the dimensionless quantity $\alpha$ introduced in
(\ref{alpha}).  Because of the analogy with Aharonov-Bohm scattering, we
expect this contribution to be of order $\alpha^3$ and not of order
$\alpha^2$.  We calculate these contributions again using the effective
Lagrangian (\ref{theory}).

To identify the relevant Feynman graphs, we apply---as is commonly done in
the context of effective field theories---dimensional analysis to the
higher-order contributions to the scattering amplitude.  Following Weinberg
\cite{Weinberg}, we assume that the energies and momenta of the two incoming
and the two outgoing phonons are small and all of the same order, $p_0 = c
P, |{\bf p}| = P$ say.  As argued in that reference, the loop integrals will
be dominated by contributions of the order $P$, so that one can make a
perturbation in powers of $P$.  More specifically, each energy and momentum
appearing in the interaction vertices contributes a factor $P$.  An internal
phonon line therefore contributes a factor $1/P^2$ [see Eq.\
(\ref{phiprop})], while an internal Chern-Simons line contributes a factor
$1/P$ [see Eq.\ (\ref{CSprop})].  An integration over a loop frequency and
loop wave vectors contributes a factor $P^3$ in two space dimensions.

\begin{figure}
\begin{center}
\epsfxsize=8.cm \mbox{\epsfbox{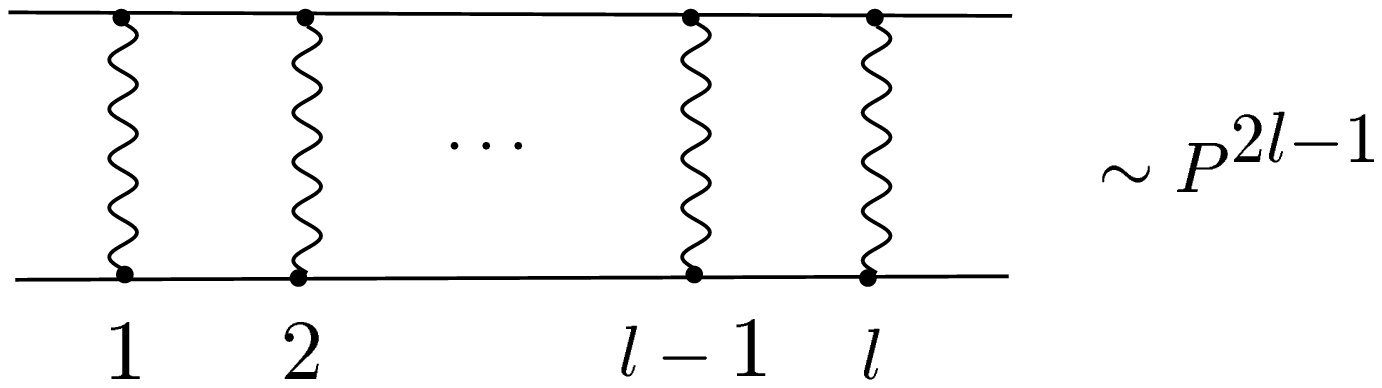}}
\end{center}
\figcap{Contribution to the elastic scattering amplitude owing to the
exchange of $l$ Chern-Simons quanta.  The incoming and outgoing phonons have
an energy and momentum of the order $cP$ and $P$, respectively.}
\label{fig:dimensional}
\end{figure}

It is now easily shown that the exchange of $l$ Chern-Simons quanta
yields a contribution of the order of $P^{2l-1}$ to the scattering
amplitude (see Fig. \ref{fig:dimensional}), 
where we also took into account the energy-dependence contained in the
normalization factor (\ref{zeta}) of the external phonon lines.  For
the exchange of a single Chern-Simons quantum ($l=1$), we recover the
momentum dependence given in (\ref{tot}).  Since the two vertices involved
are both proportional to the charge $e$ with which the phonons couple to the
Chern-Simons field describing the vortices, these diagrams are proportional
to $e^{2 l}$.  Including also the other constants contained in the vertices
and the propagators, we find that the dimensionless expansion parameter for
this type of diagrams is $\alpha^2$.  These diagrams represent the
scattering of a phonon from a vortex, and are the contributions we wish to
evaluate.

For comparison, let us consider the case where instead of Chern-Simons
quanta, phonons are exchanged.  The interaction vertices involved in these
contributions arise from the nonlinear terms in the effective theory
which we have ignored in (\ref{theory}).  Using similar arguments as the
ones above, one can readily show that the exchange of $l'$ phonons yields a
contribution to the phonon-phonon scattering amplitude of the order of
$P^{3l'-1}$.  To be more specific, the dimensionless expansion parameter is
here
\begin{equation} 
\beta = \frac{\xi}{\bar{n}} |{\bf k}|^3.
\end{equation} 
It is gratifying to see that the contributions which do not involve the
Chern-Simons field describing the vortices have a different $P$-dependence.

From the above dimensional analysis, it follows that to order
$\alpha^3$, the contributions to the scattering amplitude are of the
form
\begin{equation} \label{Cs}
A^{(1)}_{\rm tot}(\theta,|{\bf p}|) = i \frac{h \alpha}{m} \left[C_0(\theta)
+ C_1(\theta,|{\bf p}|) \alpha^2 \right],
\end{equation} 
where naively the coefficient of the second term, $C_1$, is expected to
be a function of the scattering angle $\theta$ only, in the same way as
the coefficient of the first term,
\begin{equation} \label{C0}
C_0(\theta) = \cot(\theta),
\end{equation} 
depends only on $\theta$.  However, a one-loop diagram produces a
nonanalytic term of the form $(h \alpha^3/m)\ln(|{\bf p}|)$, so that $C_1$
depends also on $|{\bf p}|$.  Whether such logarithmic terms will appear can
{\it a priori} not be determined solely by dimensional arguments because
that function is dimensionless.

The calculation of the relevant one-loop corrections to the elastic
phonon-phonon scattering amplitude is somewhat technical and therefore
relegated to the Appendix.  The final result for the coefficient
$C_1(\theta,|{\bf p}|)$ introduced in (\ref{Cs}), which we split as
\begin{equation}  
C_1(\theta,|{\bf p}|) := C_1(\theta) + C_{\rm na}(|{\bf p}|)
\end{equation} 
reads
\begin{figure}
\vspace{-4.cm}
\begin{center}
\epsfxsize=10.cm \mbox{\epsfbox{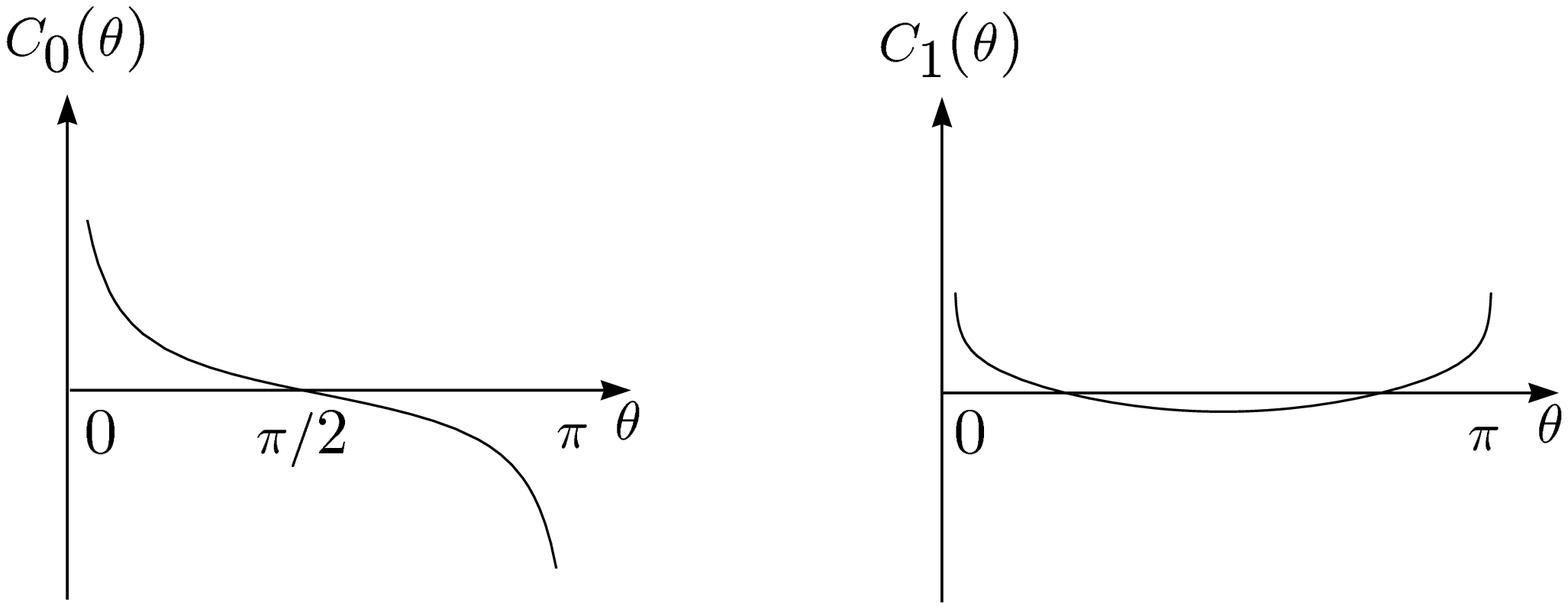}}
\end{center}
\vspace{-4.cm}
\figcap{Graphical representation of the ${\cal
O}\left(\alpha\right)$-contribution $C_0(\theta)$ and the ${\cal
O}\left(\alpha^3\right)$-contribution $C_1(\theta)$ to the elastic
phonon-phonon scattering amplitude.}
\label{fig:amplitude}
\end{figure}
\begin{eqnarray}  \label{final}
C_1(\theta) = - \frac{1}{8 \pi^2} && \hspace{-.7cm} \Biggl\{ \frac{5}{3} +
\sin(\tfrac{1}{2} \theta) \cos(\tfrac{1}{2} \theta) [\tfrac{1}{2} \pi - 2
\sin(\tfrac{1}{2} \theta) \cos(\tfrac{1}{2} \theta)]  \\ &&
\hspace{-.5cm} + \sin^3(\tfrac{1}{2} \theta) \left[ \ln \left( \frac{1 -
\sin(\tfrac{1}{2} \theta)}{\cos(\tfrac{1}{2} \theta)} \right) - \frac{1}{2}
\left( \pi + \frac{\theta - \pi}{\cos(\tfrac{1}{2} \theta)} \right) \right]
\nonumber \\ && \hspace{-.5cm} + \cos^3(\tfrac{1}{2} \theta) \left[ \ln
\left( \frac{\sin(\tfrac{1}{2} \theta)}{1 + \cos(\tfrac{1}{2} \theta)}
\right) - \frac{1}{2} \left( \pi - \frac{\theta}{\sin(\tfrac{1}{2} \theta)}
\right) \right] \Biggr\}, \nonumber 
\end{eqnarray} 
and 
\begin{equation} 
C_{\rm na}(|{\bf p}|) = -\frac{1}{16 \pi^3} \int_{0}^{1} \dd y
\frac{1}{y^2-1}.
\end{equation} 
This last contribution, which is obtained from Eq.\ (\ref{IIIdiv}) with the
crossed term included, is logarithmically diverging.  It is independent of
the scattering angle.  A closer inspection reveals that the divergence is an
infrared divergence arising when the momentum of one of the Chern-Simons
propagators in the diagram depicted in Fig.\ \ref{fig:eighteen} tends to
zero.  To regularize the integral, we give the Chern-Simons field a small
mass $\mu$; $C_{\rm na}(|{\bf p}|)$ then becomes
\begin{eqnarray} \label{na} 
C_{\rm na}(|{\bf p}|) &=& -\frac{1}{32 \pi^3} \int_{0}^{1} \dd y
\frac{y-1}{(y+1)(y-\sqrt{1-\mu^2 c^2/{\bf p}^2})^2} \nonumber \\ && +
\frac{1}{32 \pi^3} \int_{1}^{\infty} \dd y
\frac{y-1}{(y+1)(y-\sqrt{1-\mu^2 c^2/{\bf p}^2})^2} \nonumber \\
&\rightarrow& \frac{1}{16 \pi^3}\left[\ln\left(\frac{2|{\bf p}|}{\mu
c}\right) - \frac{1}{2} \right].
\end{eqnarray} 
This introduces an arbitrary scale factor into the problem.  Frequently,
such a scale factor arises instead when ultraviolet divergences are
present.  Logarithmic divergences of this kind have to be eliminated by
a redefinition of certain parameters of the effective theory at the
expense of the appearance of an arbitrary renormalization scale
\cite{Weinberg}.  The nonrelativistic effective theory considered here
is ultraviolet finite to this order, and needs, therefore, not to be
renormalized.  The arbitrary scale factor $\mu$ arises here from the
infrared region.

A graphical representation of the analytic part of the one-loop result,
$C_1(\theta)$, together with the tree result is given in
Fig. \ref{fig:amplitude}.  It shows that whereas the tree contribution
$C_0(\theta)$ is antisymmetric about $\theta=\pi/2$, the ${\cal
O}\left(\alpha^3\right)$-contributions are symmetric.  Both are seen to
diverge for $\theta=0$ and $\pi$.
\section{Discussion}
We have extended the analysis of the quantum field theory recently
proposed by these authors \cite{JA} to describe vortices in a superfluid
film at the absolute zero of temperature.  The theory consists of the
effective action of phonons coupled to a Chern-Simons term.  Two of its
salient features are that, first, the effective action describing the
phonons is invariant under Galilei transformations---as is required for
a nonrelativistic system---and, second, the coupling to the Chern-Simons
terms is linear and not minimal.  We have demonstrated that various
known facts about two-dimensional quantum vortices are correctly
reproduced by the quantum field theory.

We have further shown how it can be used to calculate one-loop
contributions to the amplitude for the elastic scattering of phonons
from a vortex.  We have applied dimensional analysis of the type
commonly used in the context of effective field theories to identify the
relevant Feynman graphs and have calculated these.  We have shown that
the one-loop corrections are ultraviolet finite, and that an arbitrary
scale factor was introduced by an infrared divergence.

By minimally coupling the Goldstone field $\varphi$ to electrodynamics,
the theory discussed here may be extended to describe vortices in a
superconducting film.
\vspace{.5cm} \\
\noindent
{\bf Acknowledgments} \\
\noindent
We wish to thank H. Kleinert for useful discussions.  This work was
performed as part of a scientific network supported by the European
Science Foundation, an association of 62 European national funding
agencies (see network's URL,
http://www.physik.fu-berlin.de/$\sim$defect).
\renewcommand{\theequation}{A.\arabic{equation}}
\setcounter{equation}{0}
\section*{Appendix}
\begin{figure}
\begin{center}
\epsfxsize=4.cm 
\mbox{\epsfbox{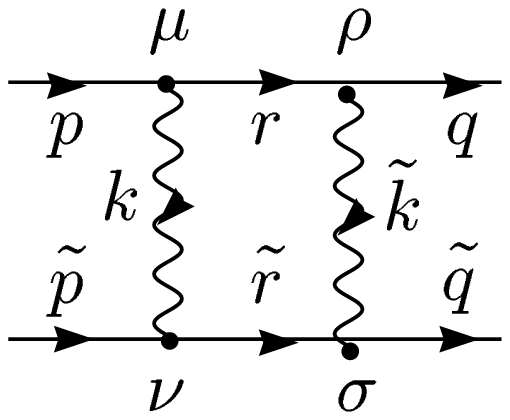}}
\end{center}
\figcap{One-loop diagram I contributing to the scattering amplitude.}
\label{fig:seventeen}
\end{figure}

\begin{figure}
\begin{center}
\epsfxsize=5.cm
\mbox{\epsfbox{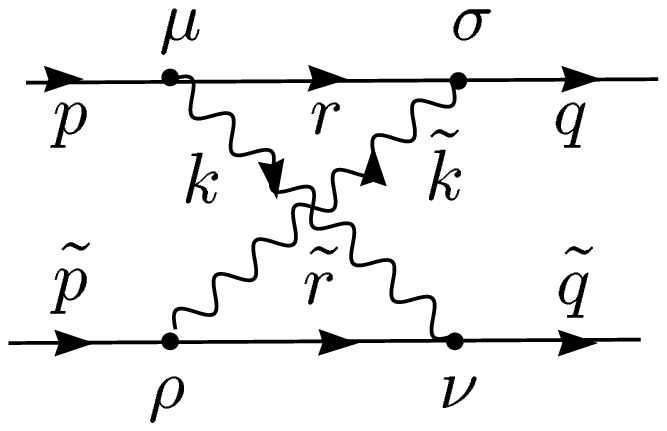}}
\end{center}
\figcap{One-loop diagram II contributing to the scattering amplitude.}
\label{fig:nineteenp}
\end{figure}

\begin{figure}
\begin{center}
\epsfxsize=5.cm
\mbox{\epsfbox{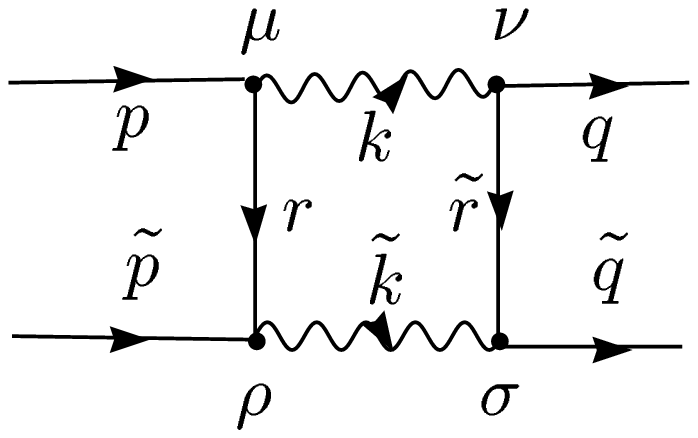}}
\end{center}
\figcap{One-loop diagram III contributing to the scattering amplitude.}
\label{fig:eighteen}
\end{figure}

In this Appendix, we calculate the contributions to the elastic
phonon-phonon scattering amplitude proportional to $\alpha^3$.  These
contributions arise from the three one-loop diagrams depicted in Figs.\
\ref{fig:seventeen}-\ref{fig:eighteen}.  Since the dimensional analysis of
the previous section already fixed the dimensionful prefactors appearing in
the scattering amplitude, we can simplify matters by setting all
dimensionful parameters to one,
\begin{equation} 
\hbar=c=m=\bar{n}=1.
\end{equation} 

The Feynman rules (\ref{CSprop})-(\ref{zwee}) applied to the diagram
depicted in Fig.\ \ref{fig:seventeen}, which we refer to as I, yield
\begin{equation} \label{A1}
\mbox{I} = \zeta^4 \int \frac{\dd^3 r}{(2 \pi)^3} G(r)
G(\tilde{r})G_{\mu\nu}(k) G_{\rho\sigma}(\tilde{k}) \Gamma_\mu(p,-r)
\Gamma_\nu(\tilde{p},-\tilde{r}) \Gamma_\rho(r,-q)
\Gamma_\sigma(\tilde{r},-\tilde{q}),
\end{equation} 
where because of conservation of energy and momentum at the vertices we have
in addition to (\ref{relations}) the following relations
\begin{equation} \label{A2}
\tilde{r}_0 = 2p_0 - r_0, \; k_0 = p_0 - r_0, \; \tilde{k}_0 = r_0 -
p_0, \, \tilde{\bf r} = -{\bf r}, \; {\bf k} = {\bf p} - {\bf r}, \;
\tilde{\bf k} ={\bf r} - {\bf q} .
\end{equation} 
The summation over the indices in (\ref{A1}) (using an Euclidean metric) gives
\begin{eqnarray} \label{NA1}
\lefteqn{G_{\mu\nu}(k) G_{\rho\sigma}(\tilde{k})\Gamma_\mu(p,-r)
\Gamma_\nu(\tilde{p},-\tilde{r}) \Gamma_\rho(r,-q)
\Gamma_\sigma(\tilde{r},-\tilde{q}) = } \nonumber \\ && - \frac{1}{{\bf
k}^2 \tilde{\bf k}^2} \left[{\bf p}\cdot{\bf r}(p_0 \, \tilde{\bf r}
\times {\bf k} -\tilde{r}_0 \, {\bf p} \times {\bf k}) + {\bf p} \cdot
\tilde{\bf r} (p_0 \, {\bf r} \times {\bf k} + r_0 \, {\bf p}\times {\bf
k})\right] \nonumber\\ && \;\;\;\;\;\;\;\;\;\, \times [{\bf q} \cdot
{\bf r} (p_0 \, \tilde{\bf r} \times \tilde{\bf k} -\tilde{r}_0 \, {\bf
q} \times \tilde{\bf k}) + {\bf q} \cdot \tilde{\bf r} (p_0 \, {\bf r}
\times \tilde{\bf k} + r_0 \, {\bf q} \times \tilde{\bf k})].
\end{eqnarray}
Substituting this in (\ref{A1}) and using the relations (\ref{A2}), we
obtain the expression
\begin{equation} \label{A3} 
\mbox{I} = 4 \int \frac{\dd^3 r}{(2\pi)^3} \frac{1}{(r_0^2 - {\bf r}^2 +
i \eta) [(2 p_0 - r_0)^2-{\bf r}^2 + i\eta]} \frac{{\bf p}\cdot{\bf r}
\, {\bf p} \times{\bf r} \, {\bf q}\cdot{\bf r} \, {\bf q} \times{\bf
r}}{({\bf p}-{\bf r})^2 ({\bf q}-{\bf r})^2}.
\end{equation}  
With the help of contour integration, the integral over the loop energy
$r_0$ is readily carried out with the result
\begin{equation}  
\mbox{I} = i \int \frac{\dd^2 r}{(2\pi)^2} \frac{{\bf p}\cdot{\bf r} \, {\bf p}
\times {\bf r} \, {\bf q} \cdot {\bf r} \, {\bf q} \times {\bf r}}{|{\bf
r}|({\bf r}^2-p_0^2)({\bf p}-{\bf r})^2({\bf q}-{\bf r})^{2}} .
\end{equation} 
The integral over the angle $\alpha$ is most easily carried out by
introducing the variable $z = \exp(i \alpha)$ and performing a contour
integration along the unit circle.  The poles one encounters are located at
\begin{equation} 
z=0, \, z=y, \, z=1/y, \, z = {\rm e}^{i \theta} y, \, z = {\rm e}^{i
\theta}/y,
\end{equation} 
where $y$ stands for 
\begin{equation} \label{y}
y = \frac{|{\bf r}|}{p_0}. 
\end{equation} 
Since only those poles lying inside the unit circle are to be included,
the regions $0 \leq y < 1$ and $1 < y < \infty$ have to be treated
separately.  Adding the two contributions, we find
\begin{equation} \label{A12}
\mbox{I} = \frac{ip_0^3}{8} \int_{0}^{1} \frac{\dd y}{2\pi} (y^2 + 1) \frac{
\cos(2\theta) + 2\cos(2\theta) y^2 - [1 + 2 \cos(\theta)] y^4}{1- 2
\cos(\theta) y^2 + y^4}.
\end{equation} 
The remaining integral over $y$ is elementary and yields
\begin{equation} 
\mbox{I} = \frac{-ip_0^3}{16\pi} \left\{\frac{10}{3} + \frac{14}{3}
\cos(\theta) + 4 \cos^3(\tfrac{1}{2} \theta)
\ln[\tan(\tfrac{1}{4}\theta)] \right\}.
\end{equation} 
This result is valid for $0 \leq \theta \leq 2 \pi$.

We continue with the evaluation of the second one-loop diagram depicted in
Fig.\ \ref{fig:nineteenp}.  For this diagram II we find
\begin{equation} \label{B1}
\mbox{II} = \zeta^4 \int \frac{\dd^3 r}{(2 \pi)^3} G(r) G(\tilde{r})
G_{\mu\nu}(k) G_{\rho\sigma}(\tilde{k}) \Gamma_\mu(p,-r)
\Gamma_\nu(\tilde{r},-\tilde{q}) \Gamma_\rho(\tilde{p},-\tilde{r})
\Gamma_\sigma(r,-q),
\end{equation} 
where now
\begin{equation} \label{B2}
\tilde{r}_0 = r_0, \, k_0 = p_0 - r_0, \, \tilde{k}_0 = p_0 - r_0, \,
\tilde{\bf r} = {\bf r} - {\bf p} -{\bf q}, \, {\bf k} = {\bf p} - {\bf
r}, \, \tilde{\bf k} = {\bf q} - {\bf r}.
\end{equation} 
Explicitly,
\begin{eqnarray} \label{NB1}
\lefteqn{G_{\mu\nu}(k) G_{\rho\sigma}(\tilde{k})\Gamma_\mu(p,-r)
\Gamma_\nu(\tilde{r},-\tilde{q}) \Gamma_\rho(\tilde{p},-\tilde{r})
\Gamma_\sigma(r,-q) = } \nonumber \\ &&  \frac{1}{{\bf
k}^2 \tilde{\bf k}^2} \left[{\bf p}\cdot{\bf r}(p_0 \, \tilde{\bf r}
\times {\bf k} -\tilde{r}_0 \, {\bf q} \times {\bf k}) + {\bf q} \cdot
\tilde{\bf r} (p_0 \, {\bf r} \times {\bf k} + r_0 \, {\bf p}\times {\bf
k})\right] \nonumber\\ && \;\;\;\;\;\;\, \times [{\bf p} \cdot
\tilde{\bf r} (p_0 \, {\bf r} \times \tilde{\bf k} + r_0 \, {\bf
q} \times \tilde{\bf k}) + {\bf q} \cdot {\bf r} (p_0 \, \tilde{\bf r}
\times \tilde{\bf k} - \tilde{r}_0 \, {\bf p} \times \tilde{\bf k})].
\end{eqnarray}
Substituting this in (\ref{B1}) and using the relations (\ref{B2}), we
arrive at
\begin{eqnarray} \label{B4}
\mbox{II} &=& - \frac{1}{4 p_0^2} \int \frac{\dd^3 r}{(2\pi)^3}
\frac{(p_0 + r_0)^2}{(r_0^2 - {\bf r}^2 + i\eta)[r_0^2 -({\bf r}-{\bf
p}-{\bf q})^2 + i\eta]} \frac{1}{({\bf p} - {\bf r})^2 ({\bf q} - {\bf
r})^2} \nonumber \\ && \;\;\;\;\;\;\;\;\;\;\;\;\;\;\;\;\;\;\;\; \times
[{\bf p} \cdot {\bf r} ({\bf p}\times{\bf q} + {\bf q} \times {\bf
r})-{\bf q} \cdot ({\bf r} - {\bf p} - {\bf q}) {\bf p} \times{\bf r} ]
\nonumber \\ && \;\;\;\;\;\;\;\;\;\;\;\;\;\;\;\;\;\;\;\; \times [-{\bf
q} \cdot {\bf r}(-{\bf p} \times {\bf q} +{\bf p} \times {\bf r}) +{\bf
p} \cdot ({\bf r}-{\bf p}-{\bf q}) {\bf q} \times {\bf r}].
\end{eqnarray} 
Let us first carry out the integral over the loop energy $r_0$:
\begin{equation} \label{NB3}
\int \frac{\dd r_0}{2\pi} \frac{(p_0 + r_0)^2}{(r_0^2 - {\bf r}^2 + i
\eta)(r_0^2 - {\bf v}^2 + i\eta)} = -\frac{i}{2} \frac{(p_0 -
{\bf r})^2}{|{\bf r}|({\bf r}^2 - {\bf v}^2)} -\frac{i}{2}\frac{(p_0 -
|{\bf v}|)^2}{|{\bf v}|({\bf v}^2 - {\bf r}^2)},
\end{equation} 
where we introduced the abbreviation ${\bf v} = {\bf r}-{\bf p}-{\bf
q}$.  The first term in (\ref{NB3}) gives a contribution II$_1$ to II
and the second term a contribution II$_2$, such that $\mbox{II} =
\mbox{II}_1 + \mbox{II}_2$. It turns out that both contributions are
identical, so that $\mbox{II} = 2 \mbox{II}_1$.  We perform the angle
integration again by introducing the variable $z = \exp(i \alpha)$ and
carrying out a contour integration along the unit circle.  Besides poles
located at $z = y, {\rm e}^{i \theta} y$ and $z = 1/y, {\rm e}^{i
\theta}/ y$, with $y$ defined in (\ref{y}), there are also two poles at
\begin{equation}  
z_{1,2} = \frac{1 + {\rm e}^{i\theta} \pm \sqrt{({\rm
e}^{i\theta}+1)^2 - 4{\rm e}^{i\theta} y^2}}{2y} .
\end{equation} 
The contributions of the first two pairs cancel, so that we only have to
consider the contributions stemming from the poles $z_1$ and $z_2$.  For
$y \geq \cos(\tfrac{1}{2} \theta)$ these poles lie precisely on the unit
circle, $|z_{1,2}| =1$ and it is {\it a priori} not clear how to
proceed.  A closer (numerical) inspection reveals that the contributions
from $z_1$ and $z_2$ cancel there.  This leaves us with region $y <
\cos(\tfrac{1}{2} \theta)$, where $|z_1| <1$ and $|z_2| > 1$. In this
way, we obtain with $t=\cos(\tfrac{1}{2} \theta)$ the result
\begin{eqnarray} \label{B17}
\mbox{II} &=& \frac{-i p_0^3}{16 \pi} \int_0^t \dd y \,
\frac{4t(1-y)[y^2 (t^2 - 1) + t^2]}{(1+y) \sqrt{t^2 - y^2}}  
\nonumber \\ &=& \frac{-ip_0^3}{16\pi}\Bigl\{-4\sin^2( \tfrac{1}{2} \theta)
\cos^2(\tfrac{1}{2} \theta) + \frac{\pi}{2} \cos(\tfrac{1}{2} \theta) [4 -
5\cos^2(\tfrac{1}{2} \theta) - \cos^4 (\tfrac{1}{2}\theta)] \nonumber \\ &&
\;\;\;\;\;\;\;\;\;\;\; + 2 \theta \cot(\tfrac{1}{2}\theta)
\cos(\theta)\Bigr\} ,
\end{eqnarray} 
which is valid for $-\pi \leq \theta \leq  \pi$.

The last one-loop contribution to the phonon-phonon scattering amplitude
is depicted in Fig.\ \ref{fig:eighteen}.  It stands for
\begin{equation} \label{C1}
\mbox{III} = \zeta^4 \int \frac{\dd^3 r}{(2 \pi)^3} G(r)
G(\tilde{r}) G_{\mu\nu}(k) G_{\rho\sigma}(\tilde{k}) \Gamma_\mu(p,-r)
\Gamma_\nu(-q,-\tilde{r}) \Gamma_\rho(\tilde{p},r)
\Gamma_\sigma(-\tilde{q},\tilde{r}),
\end{equation}
where now we have the relations 
\begin{equation} \label{C2}
\tilde{r}_{0}=-r_{0} , \, k_0 = p_0 - r_0 , \, \tilde{k}_0 = p_0 + r_0 , \,
\tilde{\bf r} = {\bf p} - {\bf q} -{\bf r} , \, {\bf k} = {\bf p} - {\bf r},
\, \tilde{\bf k} = {\bf r} - {\bf p}. 
\end{equation} 
More explicitly,
\begin{eqnarray} \label{NC1} 
\lefteqn{G_{\mu\nu}(k) G_{\rho\sigma}(\tilde{k}) \Gamma_\mu(p,-r)
\Gamma_\nu(-q,-\tilde{r}) \Gamma_\rho(\tilde{p},r)
\Gamma_\sigma(-\tilde{q},\tilde{r}) = } \nonumber \\ && \frac{1}{{\bf k}^2
\tilde{\bf k}^2} [{\bf p} \cdot {\bf r}(p_0 \, \tilde{\bf r} \times {\bf k}
+ \tilde{r}_0 \, {\bf q} \times {\bf k}) - {\bf q} \cdot \tilde{\bf r} (p_0
\, {\bf r} \times{\bf k} + r_0 \, {\bf p} \times{\bf k})] \nonumber \\ &&
\;\;\;\;\;\; \times [{\bf p}\cdot {\bf r}(p_0 \, \tilde{\bf r} \times
\tilde{\bf k} - \tilde{r}_0 \, {\bf q} \times \tilde{\bf k}) - {\bf q} \cdot
\tilde{\bf r} (p_0 \, {\bf r} \times \tilde{\bf k} - r_0 \, {\bf p} \times
\tilde{\bf k})].
\end{eqnarray} 
In this way, (\ref{C1}) becomes
\begin{eqnarray} \label{C3}
\mbox{III} &=& \frac{-1}{4p_0^2} \int \frac{\dd^3 r}{(2\pi)^3} \frac{p_0^2 -
r_0^2}{(r_0^2 - {\bf r}^2 + i\eta) [r_0^2 - ({\bf p} - {\bf q} - {\bf r})^2
+ i\eta]} \\ && \;\;\;\;\;\;\;\;\;\;\;\;\;\;\;\;\;\; \times \frac{[{\bf p}
\cdot {\bf r} ({\bf p} \times {\bf q} + {\bf q} \times {\bf r}) + {\bf q}
\cdot ({\bf p}-{\bf q}-{\bf r}) {\bf p} \times {\bf r}]^2}{({\bf p} - {\bf
r})^4} . \nonumber
\end{eqnarray} 
We again first carry out the integral over the loop energy $r_0$,
\begin{equation} \label{NC3}
 \int \frac{\dd r_0}{2\pi} \frac{p_0^2 - r_0^2}{(r_0^2 - {\bf r}^2 +
i\eta) (r_0^2 - {\bf w}^2 + i\eta )} = 
\frac{i}{2} \frac{{\bf r}^2 - p_0^2}{|{\bf r}|({\bf r}^2 - {\bf w}^2)} +
\frac{i}{2} \frac{{\bf w}^2 - p_0^2}{|{\bf w}|({\bf w}^2 - {\bf r}^2)},
\end{equation} 
where ${\bf w} = {\bf r}+{\bf q}-{\bf p}$.  The first (second) term
gives a contribution $\mbox{III}_{1(2)}$, such that $\mbox{III} =
\mbox{III}_1 + \mbox{III}_2$.  The angle integral is again evaluated by
a contour integration along the unit circle.  With $y$ as defined in
(\ref{y}), the integrand of $\mbox{III}_1$ has poles at $z= y,1/y$ as
well as at
\begin{equation} 
z_{3,4} = \frac{1-{\rm e}^{i\theta} \mp \sqrt{({\rm
e}^{i\theta}-1)^2 + 4{\rm e}^{i\theta} y^2}}{2y} ,
\end{equation} 
while the integrand of $\mbox{III}_2$ has poles at $z= {\rm e}^{i \theta}
y,{\rm e}^{i \theta}/y$ and at
\begin{equation} 
z_{5,6} = \frac{-1 + {\rm e}^{i\theta} \mp \sqrt{({\rm
e}^{i\theta}-1)^2 + 4{\rm e}^{i\theta} y^2}}{2y} .
\end{equation}
In $\mbox{III}_2$, we made the change of integration variable ${\bf r}
\rightarrow {\bf w}$.  The contributions from the poles $z=y,1/y,{\rm e}^{i
\theta} y,{\rm e}^{i \theta}/y$ can be calculated in a straightforward
manner.  They cause a logarithmic-diverging integral
\begin{equation} \label{IIIdiv}
\mbox{III}_{\rm na} = \frac{i p_0^3}{4 \pi} \sin^2(\tfrac{1}{2}\theta)
\int_{0}^{1} \dd y \frac{1}{y^2-1},
\end{equation} 
which is discussed in the main text.  The remaining poles $z_3,z_4,z_5$, and
$z_6$ which we have to consider lie precisely on the unit circle for $y \geq
\sin(\tfrac{1}{2} \theta)$.  A careful numerical analysis reveals that their
contributions cancel here.  This leaves us with the region $y \leq
\sin(\tfrac{1}{2} \theta)$, where only the poles $z_3$ and $z_6$ have to be
included since $|z_{3,6}| < 1$ while $|z_{4,5}| > 1$.  Both $z_3$ and $z_6$
give the same real part, while their imaginary parts cancel.  With $s$
denoting $\sin(\tfrac{1}{2} \theta)$, we finally obtain for diagram III
\begin{eqnarray} \label{C17}
\lefteqn{\mbox{III} - \mbox{III}_{\rm na} = } \\ && \frac{i p_0^3}{16
\pi} \int_0^s \dd y \, \frac{4s(y^2 - 1)[-s^2 + y^2 + (1-s^2)
y^4]}{\sqrt{s^2 - y^2}\left[y^4 - 2s y^2 \left(s - \sqrt{s^2 -
y^2}\right) + \left(s - \sqrt{s^2 - y^2}\right)^2\right]^2} \nonumber \\
&& \;\;\;\;\;\;\;\;\;\;\;\;\;\;\;\; \times \left(8s^4 - 8 s^2 y^2 + y^4
- 8 s^3 \sqrt{s^2 - y^2} + 4 s y^2 \sqrt{s^2 - y^2}\right) = \nonumber
\\ && \frac{-i p_0^3}{16 \pi} \left\{ 2\pi |\sin(\tfrac{1}{2}\theta)|
\cos(\tfrac{1}{2} \theta) - \frac{\pi}{2} |\sin(\tfrac{1}{2} \theta)|
\left[ 4-\sin^2(\tfrac{1}{2}\theta) - \sin^4(\tfrac{1}{2}\theta) \right]
\right\} , \nonumber
\end{eqnarray} 
where $\theta$ is to be restricted to the values $-\pi \leq \theta \leq
\pi$.

Since we are dealing with identical particles, we have to include the
crossed diagrams whose contributions are obtained from the uncrossed
diagrams by replacing $\theta \rightarrow \theta - \pi \; \mbox{mod}(2
\pi)$.  More specifically, restricting ourselves to the values $0 \leq
\theta \leq \pi$, we have to replace $\theta$ with $\theta + \pi$ in the
result (\ref{A12}) of diagram I, and with $\theta - \pi$ in the results
(\ref{B17}) and (\ref{C17}) of diagram II and III.  The final expression for
the one-loop contributions to the elastic phonon-phonon scattering
proportional to $\alpha^3$ is given in (\ref{final}) and (\ref{na}).
\end{document}